\documentclass[journal,twocolumn,10pt]{IEEEtran}
\usepackage[T1]{fontenc}% optional T1 font encoding
\usepackage{cite}
\usepackage[cmex10]{amsmath}
\usepackage{amsthm}
\usepackage{amssymb}
\usepackage{mathtools}
\usepackage[hidelinks]{hyperref} 
\usepackage{algorithmic}
\usepackage{textcomp}
\usepackage[usenames, dvipsnames]{color}
\usepackage[linesnumbered,ruled]{algorithm2e}
\usepackage{subcaption}
\usepackage{graphicx}  % remove 'demo' option for your real document
\usepackage{url}
\usepackage[usestackEOL]{stackengine}
\usepackage{verbatim}

\usepackage {tikz}
\tikzset{>=latex}
\usepackage{xcolor,colortbl}
\definecolor{LayerColor}{RGB}{230, 230, 230}
\definecolor{InOutColor}{RGB}{240, 243, 255}
\definecolor{cellColor}{RGB}{230, 230, 230}

\newcommand{\mc}[1]{\mathcal{#1}}

\newcommand{\mb}[1]{\mathbf{#1}}
\newcommand{\mr}[1]{\mathrm{#1}}

\DeclareMathOperator*{\argmin}{arg\;min}

\ifCLASSINFOpdf
\else
\fi
\usepackage{amsmath}
\begin{document}
\title{Deep Learning for Estimation and Pilot Signal Design in Few-Bit Massive MIMO Systems}

\author{Ly~V.~Nguyen,
	 Duy~H.~N.~Nguyen, and A.~Lee~Swindlehurst
	\thanks{Ly V. Nguyen is with the Computational Science Research Center, San Diego State University, San Diego, CA, USA 92182 (e-mail: vnguyen6@sdsu.edu).}
	\thanks{Duy H. N. Nguyen is with the Department of Electrical and Computer Engineering, San Diego State University, San Diego, CA, USA 92182 (e-mail: duy.nguyen@sdsu.edu).}
	\thanks{A. Lee Swindlehurst is with the Center for Pervasive Communications and Computing, Henry Samueli School of Engineering, University of California, Irvine, CA, USA 92697 (e-mail: swindle@uci.edu).}
}

\maketitle

\begin{abstract}
%	Low-resolution analog-to-digital converters (ADCs) have been suggested as an efficient and practical solution to reducing hardware cost and power consumption in massive MIMO systems. Unfortunately, the 
Estimation in few-bit MIMO systems is challenging, since the received signals are nonlinearly distorted by the low-resolution ADCs. In this paper, we propose a deep learning framework for channel estimation, data detection, and pilot signal design to address the nonlinearity in such systems. The proposed channel estimation and data detection networks are model-driven and have special structures that take advantage of the domain knowledge in the few-bit quantization process. While the first data detection network, namely B-DetNet, is based on a linearized model obtained from the Bussgang decomposition, the channel estimation network and the second data detection network, namely FBM-CENet and FBM-DetNet respectively, rely on the original quantized system model. To develop FBM-CENet and FBM-DetNet, the  maximum-likelihood channel estimation and data detection problems are reformulated to overcome the vanishing gradient issue. An important feature of the proposed FBM-CENet structure is that the pilot matrix is integrated into its weight matrices of the channel estimator. Thus, training the proposed FBM-CENet enables a joint optimization of both the channel estimator at the base station and the pilot signal transmitted from the users. Simulation results show significant performance gain in estimation accuracy by the proposed deep learning framework.
\end{abstract}


\setlength\arraycolsep{2pt}

% \fi
\IEEEpeerreviewmaketitle

\section{Introduction}
\label{sec_introduction}
%Ubiquitous access to high speed wireless links is a critical requirement for 5G-and-beyond networks in order to deal with the proliferation of mobile devices and services~\cite{Boccardi2014Five,andrews2014will}. Massive multiple-input multiple-output (MIMO) is one of the core technologies to satisfy such a requirement thanks to its ability of dramatically improving capacity and spectral efficiency~\cite{Hoydis2013massive,ngo2013energy,Lu2014Overview}. Massive MIMO uses a large number (tens to hundreds) of antennas at the base station in order to exploit the large spatial degrees of freedom. However, the use of a large number of antennas at the base station puts a heavy burden on hardware cost and power consumption since many radio-frequency (RF) chains are required. %A conventional RF chain consists of a series of components such as band-pass filter (BPF), low-noise amplifier (LNA), mixer, low-pass filter (LPF), automatic gain control (AGC), and high-precision analog-to-digital converters (ADC). Scaling such an implementation to many antennas in massive MIMO systems can be too expensive and power-consuming.

One practical solution for reducing hardware cost and power consumption in massive MIMO systems is to use low-resolution (e.g., $1$--$3$ bits) analog-to-digital convectors (ADCs). This is due to the simple structure and very low power consumption of low-resolution ADCs. In particular, the number of comparators in a $b$-bit ADC grows exponentially with $b$, which means both the hardware complexity and the power consumption of an ADC scales exponentially with the resolution \cite{walden1999analog}. Therefore, the cost and power consumption of low-resolution ADCs are substantially lower than those of high-resolution ADCs. Furthermore, the hardware structure of other components in an RF chain can also be simplified or removed when low-resolution ADCs are used. For example, the simplest architecture involving one-bit ADCs does not require an automatic gain control (AGC) since only the \textit{sign} of the real and imaginary parts of the received signals is retained. The low-noise amplifier (LNA) with a stringent requirement on linear behavior can be replaced by an amplifier whose structure is much more simpler. %Therefore, low-resolution ADCs are a promising solution for reducing cost and power consumption in massive MIMO systems. 
%Unfortunately, few-bit MIMO systems are severely nonlinear since the received signals are significantly distorted. The resulting channel estimation and data detection problems are much more challenging compared to those in unquantized systems.
Unfortunately, the nonlinearity caused by low-resolution ADCs make channel estimation and data detection in few-bit MIMO systems much more challenging, compared to those in unquantized systems.

Channel estimation for massive MIMO systems with low-resolution ADCs has attracted significant research interest and also been studied intensively. The majority of which focus on one-bit systems with different scenarios, e.g.,~\cite{choi2016near,li2017channel,Shilpa2019Massive,Zhichao2019Oversampling,Zhichao2019Channel,Liu2020Angular,Kim2018Dominant,Kim2018Channel,kim2019channel2,Srinivas2019Itervative,Mezghani2018Blind,kim2019channel,Mo2018Channel,Rodriguez2016Channel,Rusu2015Adaptive,Rao2019Channel}. Specifically, a one-bit maximum-likelihood (ML) channel estimator was proposed in \cite{choi2016near}. The work in~\cite{li2017channel} exploits the Bussgang decomposition to form a one-bit Bussgang-based minimum mean-squared error (BMMSE) channel estimator. Another BMMSE channel estimator was also proposed in~\cite{Shilpa2019Massive} but for one-bit spatial sigma-delta ADCs in a spatially oversampled array. Channel estimation with temporally oversampled one-bit ADCs is studied in~\cite{Zhichao2019Oversampling} and~\cite{Zhichao2019Channel}. It has been shown that one-bit ADCs with spatial and temporal oversampling can help improve the channel estimation accuracy but more resources and computations are required due to the oversampling process. Angular-domain channel estimation for one-bit massive MIMO systems was studied in~\cite{Liu2020Angular,Kim2018Dominant,Kim2018Channel}. Spatially/temporally correlated channels and multi-cell processing with pilot contamination were investigated in~\cite{kim2019channel2} and~\cite{Srinivas2019Itervative}, respectively. For sparse millimeter-wave MIMO channels, ML and maximum a posteriori (MAP) channel estimations were examined in~\cite{Mezghani2018Blind} and~\cite{kim2019channel}, respectively. Taking into account the sparsity of such channels, the one-bit ADC channel estimation problem has been formulated as a compressed sensing problem in~\cite{Mo2018Channel,Rodriguez2016Channel,Rusu2015Adaptive}. Performance bounds on the channel estimation of mmWave one-bit massive MIMO channels were reported in~\cite{Rao2019Channel}. 

Recently, machine learning techniques have been studied to addressing the one-bit massive MIMO channel estimation problem~\cite{Ly2021SVM,balevi2019two,Dong2020Channel,Zhang2020Deep}. The work in~\cite{Ly2021SVM} shows that support-vector machine (SVM) can be used to efficiently address the one-bit massive MIMO channel estimation problem. %The SVM-based channel estimators in~\cite{Ly2021SVM} have been shown to significantly outperform other existing one-bit channel estimators. 
Deep neural networks (DNNs) have also been used to form one-bit massive MIMO channel estimators~\cite{balevi2019two,Dong2020Channel,Zhang2020Deep}. A two-stage channel estimator for OFDM systems was proposed in~\cite{balevi2019two}. %While the work in~\cite{Dong2020Channel} considers the ray-tracing channel model, the system model in~\cite{Zhang2020Deep} assumes only one single-antenna user.
Since the majority of work in the literature focused on one-bit systems, there were limited results on few-bit massive MIMO channel estimation~\cite{Kolomvakis2020Quantized,DuyNguyen2020Neural,Gao2019Deep,Zicheng2021Deep}. Specifically, the Bussgang decomposition was exploited in~\cite{Kolomvakis2020Quantized} to derive two linear channel estimators including BMMSE and Bussgang-based weighted zero-forcing (BWZF). A DNN-based joint pilot signal and channel estimator design is proposed in \cite{DuyNguyen2020Neural}. The work in \cite{Gao2019Deep,Zicheng2021Deep} studied mixed-resolution channel estimation where low-resolution ADCs were used in only part of the receive antennas and the rest are equipped with conventional ADCs.

Data detection for low-resolution massive MIMO systems has also been studied intensively in the literature. Most of the results were reported for the case of one-bit ADCs, e.g.,~\cite{choi2016near,Jeon2018One,Kim2020Machine,Lan2018Linearized,Ly2021Linear,Demir2020ADMM,Mirfarshbafan2020Algorithm,jeon2019robust,Song2019CRC-Aided,Cho2019OneBitSCSO,Shao2018Iterative}. In particular, a one-bit ML detector and a one-bit sphere decoding (OSD) technique were proposed in~\cite{choi2016near} and~\cite{Jeon2018One}, respectively. The very high computational complexity of the ML and OSD methods nevertheless make them impractical for large-scale systems. A near-ML (nML) data detection method for large-scale MIMO systems was proposed in~\cite{choi2016near}. However, the nML method is non-robust at high signal-to-noise ratios (SNRs) when the channel state information (CSI) is not perfectly known. The learning-based method in~\cite{Kim2020Machine} is a blind detection method for which CSI is not required, but it is only applicable to MIMO systems with a small number of transmit antennas and only low-dimensional constellations. Various one-bit linear detectors were introduced in~\cite{Lan2018Linearized,Ly2021Linear}. These linear detectors are applicable for large-scale systems but often suffer from high detection error floors. The authors in~\cite{Demir2020ADMM} proposed a one-bit detection method based on the alternating direction method of multipliers (ADMM) algorithm that takes hardware impairments into account. SVM-based and DNN-based one-bit detectors were proposed in~\cite{Ly2021SVM} and~\cite{Ly2021Linear}, respectively. The SVM-based and DNN-based detectors in~\cite{Ly2021SVM} and \cite{Ly2021Linear} were shown to be robust, applicable to highly-scaled systems, and also to outperform other existing one-bit detectors. Several other one-bit data detection approaches can be found in~\cite{jeon2019robust,Song2019CRC-Aided,Cho2019OneBitSCSO,Shao2018Iterative}, but they are only applicable in systems where either a cyclic redundancy check (CRC)~\cite{jeon2019robust,Song2019CRC-Aided,Cho2019OneBitSCSO} or an error correcting code such as a low-density parity-check (LDPC) code~\cite{Shao2018Iterative} is available.

Data detection in few-bit massive MIMO systems has been studied in recent papers~\cite{wen2016bayes,Thoota2021Variational,Jeon2018supervised,Ly2020Supervised,Kolomvakis2020Quantized}. While generalized approximate message passing (GAMP) and Bayes inference are exploited in~\cite{wen2016bayes}, the work in~\cite{Thoota2021Variational} employed variational Bayesian (VB) inference and belief propagation (BP) for soft symbol decoding. However, the resulting methods can be sophisticated and expensive to implement. Unlike the blind detection method in~\cite{Kim2020Machine} which was developed for one-bit systems, the learning-based blind detection methods in~\cite{Jeon2018supervised,Ly2020Supervised} are applicable for few-bit systems, but they are also restricted to MIMO systems with a small number of transmit antennas and only low-dimensional constellations. The BMMSE and BWZF detection methods in~\cite{Kolomvakis2020Quantized} are linear detectors and thus simple and applicable for large-scale MIMO systems.

In this paper, we develop a deep learning framework for channel estimation and data detection for massive MIMO systems with low-resolution ADCs. %We consider a more general channel model which is not assumed to be sparse without any oversampling. 
Based on deep unfolding of first-order optimization iterations, we propose a channel estimator and two data detectors that are applicable for both one-bit and few-bit ADCs as well as large-scale systems without the need for CRC or error correcting codes. We note that the proposed channel estimation and data detection networks are model-driven and have special structures that can take advantages of the domain knowledge in few-bit MIMO systems. %The contributions of this paper are summarized as follows.

For channel estimation, we reformulate the ML channel estimation problem by exploiting the approximation of the cumulative distribution function (cdf) of the normal random variable as a Sigmoid activation function. The reformulated channel estimation problem does not suffer from the vanishing gradient issue as the original problem. Based on the reformulated problem and a deep unfolding technique, we propose a \underline{F}ew-\underline{B}it massive \underline{M}IMO \underline{C}hannel \underline{E}stimation \underline{Net}work, which is referred to as FBM-CENet. An interesting feature of the proposed FBM-CENet is that the pilot signal matrix is directly integrated in the weight matrices at the estimation network. When the pilot matrix is not given, it can be treated as trainable parameters and therefore training the proposed FBM-CENet is equivalent to \textit{jointly optimizing both the channel estimator at the base station and the pilot signal transmitted from the users}. This is a significant advantage of the proposed FBM-CENet structure since existing channel estimators are often designed for a known pilot matrix. Simulation results show that the proposed FBM-CENet significantly outperforms existing channel estimation methods.

For data detection, we first propose a \underline{B}ussgang-based few-bit massive MIMO Data \underline{Det}ection \underline{Net}work, referred to as B-DetNet. The proposed B-DetNet is based on a linearized system model obtained through the Bussgang decomposition.
Then we propose a \underline{F}ew-\underline{B}it massive \underline{M}IMO Data \underline{Det}ection \underline{Net}work, referred to as FBM-DetNet. Unlike B-DetNet which relies on an approximated linearized system model, FBM-DetNet is developed based on the original quantized system model. The special structure of FBM-DetNet is also obtained through a reformulated ML data detection problem whose formulation is similar to that of the reformulated channel estimation problem. We stress that the proposed B-DetNet and FBM-DetNet are highly adaptive to the channel since the weight matrices and the bias vectors of the proposed detection networks are defined by the channel matrix and the received signal vector, respectively. This makes the proposed detection networks easy to train with a few trainable parameters. Simulation results also show that the proposed data detection networks significantly outperform existing data detection methods.

The rest of this paper is organized as follows: Section~\ref{sec_system_model}
introduces the assumed system model. Channel estimation is considered in Section~\ref{sec_channel_estimation}, where the FBM-CENet estimator is proposed. The two proposed data detection networks B-DetNet and FBM-DetNet are presented in Section~\ref{sec_data_detection}. Numerical results are given in Section~\ref{sec_numerical_results}. Finally, Section~\ref{sec_conclusion} concludes the paper.

\textit{Notation:} Upper-case and lower-case boldface letters denote matrices and column vectors, respectively. $\mathbb{E}[\cdot]$ represents expectation. The operator $|\cdot|$ denotes the absolute value of a number and the operator $\|\cdot\|$ denotes the $\ell_2$-norm of a vector. The transpose is denoted by $[\cdot]^T$. The notation $\Re\{\cdot\}$ and $\Im\{\cdot\}$ respectively denotes the real and imaginary parts of the complex argument. If $\Re\{\cdot\}$ and $\Im\{\cdot\}$ are applied to a matrix or vector, they are applied separately to every element of that matrix or vector. The operator $\operatorname{vec}(\mathbf{A})$ vectorizes $\mathbf{A}$ by stacking the columns of $\mathbf{A}$ on top of one another. $\otimes$ denotes the Kronecker product. $\mathbb{R}$ and $\mathbb{C}$ denote the set of real and complex numbers, respectively, and $j$ is the unit imaginary number satisfying $j^2=-1$. $\mathcal{N}(\cdot,\cdot)$ and $\mathcal{CN}(\cdot,\cdot)$ represent the real and the complex normal distributions respectively, where the first argument is the mean and the second argument is the variance or the covariance matrix. The functions $\Phi(t) = \int_{-\infty}^{t}\frac{1}{\sqrt{2\pi}}e^{-\frac{\tau^2}{2}}d\tau$ and $\phi(t) = \frac{1}{\sqrt{2\pi}}e^{-\frac{1}{2}t^2}$ are the cdf and pdf of the standard normal random variable, respectively.

\section{System Model}
\label{sec_system_model}
We consider an uplink massive MIMO system with $K$ single-antenna users and an $N$-antenna base station, where it is assumed that $N \geq K$. Let $\bar{\mathbf{x}} = [\bar{x}_1, \bar{x}_2, \ldots, \bar{x}_K]^T \in \mathbb{C}^K$ denote the transmitted signal vector, where $\bar{x}_k$ is the signal transmitted from the $k^{\text{th}}$ user under the power constraint $\mathbb{E}[|\bar{x}_k|^2]=1$. The signal $\bar{x}_k$ is drawn from a constellation $\bar{\mathcal{M}}$. Let $\bar{\mathbf{H}} \in \mathbb{C}^{N\times K}$ denote the channel, which is assumed to be block flat fading. Let $\bar{\mathbf{r}} = [\bar{r}_1, \bar{r}_2, \ldots, \bar{r}_N]^T \in \mathbb{C}^N$ be the unquantized received signal vector at the base station, which is given as 
\begin{equation}
\bar{\mathbf{r}} = \bar{\mathbf{H}}\bar{\mathbf{x}}+\bar{\mathbf{z}}
\label{eq_analog_complex_received_signal}
\end{equation}
where $\bar{\mathbf{z}} = [\bar{z}_1, \bar{z}_2, \ldots,\bar{z}_N]^T \in \mathbb{C}^{N}$ is a noise vector whose elements are assumed to be independent and identically distributed (i.i.d.) as $\mathcal{CN}(0,N_0)$ with $N_0$ being the noise power. Each analog received signal is then quantized by a pair of $b$-bit ADCs. Hence, the quantized received signal is given by
\begin{equation}
\bar{\mathbf{y}} =  \mathcal{Q}_b\left(\Re\{\bar{\mathbf{r}}\}\right) + j\mathcal{Q}_b\left(\Im\{\bar{\mathbf{r}}\}\right).
\label{eq_quantized_complex_received_signal}
\end{equation}
The operator $\mathcal{Q}_b(\cdot)$ of a matrix or vector is applied separately to every element of that matrix or vector. The SNR is defined as $\rho = 1/N_0$. 

It is assumed that that ADCs perform $b$-bit uniform scalar quantization. The $b$-bit ADC model is characterized by a set of $2^b-1$ thresholds denoted
as $\{\tau_1,\ldots,\tau_{2^b-1}\}$. Without loss of generality, we can assume
$-\infty = \tau_0 < \tau_1 <\ldots< \tau_{2^b-1} < \tau_{2^b} = \infty$. Let $\Delta$ be the step
size, so the threshold of a uniform quantizer is given as
\begin{equation}
    \tau_l = (-2^{b-1}+l)\Delta, \; \text{for}\; l \in \mathcal{L}=\{1,\ldots,2^b-1\}.
\end{equation}
The quantization output is defined as
\begin{equation}
    \mathcal{Q}_b(r) = 
    \begin{cases}
    \tau_{l} - \frac{\Delta}{2} & \text{if}\; r\in(\tau_{l-1},\tau_l]\;\text{with}\;l\in\mathcal{L}\\
    (2^b-1)\frac{\Delta}{2}&\text{if}\;r\in(\tau_{2^b-1},\tau_{2^b}].
    \end{cases}
\end{equation}

\section{Channel Estimation in Few-Bit MIMO Systems}
\label{sec_channel_estimation}
In order to estimate the channel, a pilot sequence $\bar{\mathbf{X}}_{\mathrm{t}}\in \mathbb{C}^{K\times T_\mathrm{t}}$ of length $T_\mathrm{t}$ is used to generate the training data
\begin{equation}
\bar{\mathbf{Y}}_{\mathrm{t}} = \mathcal{Q}_b\left(\bar{\mathbf{H}}\bar{\mathbf{X}}_{\mathrm{t}}+\bar{\mathbf{Z}}_{\mathrm{t}}\right).
\label{eq_chan_est_complex_mat_form}
\end{equation}
The subscript `$\mathrm{t}$' in this paper indicates the training phase where the channel estimation task is performed. We vectorize the received signal in~\eqref{eq_chan_est_complex_mat_form} to obtain the following form:
\begin{equation}
    \bar{\mathbf{y}}_\mathrm{t} = \mathcal{Q}_b(\bar{\mathbf{P}}\bar{\mathbf{h}} + \bar{\mathbf{z}}_\mathrm{t})
    \label{eq_chan_est_complex_vec_form}
\end{equation}
where $\bar{\mathbf{y}}_\mathrm{t} = \operatorname{vec}(\bar{\mathbf{Y}}_\mathrm{t})$, $\bar{\mathbf{P}} = \bar{\mathbf{X}}_\mathrm{t}^T\otimes\mathbf{I}_{N}$, $\bar{\mathbf{h}} = \operatorname{vec}(\bar{\mathbf{H}})$, and $\bar{\mathbf{z}}_\mathrm{t} = \operatorname{vec}(\bar{\mathbf{Z}}_\mathrm{t})$. For convenience in later derivations, we convert the notation in~\eqref{eq_chan_est_complex_vec_form} into the real domain as
\begin{equation}
    \mathbf{y}_\mathrm{t} = \mathcal{Q}_b(\mathbf{P}\mathbf{h} + \mathbf{z}_\mathrm{t})
    \label{eq_chan_est_real_vec_form}
\end{equation}
where 
\begin{align*}
    \mathbf{y}_\mathrm{t} &= \begin{bmatrix}
\Re\{\bar{\mathbf{y}}_\mathrm{t}\}\\\Im\{\bar{\mathbf{y}}_\mathrm{t}\}
\end{bmatrix}, \; \mathbf{h} = \begin{bmatrix}
\Re\{\bar{\mathbf{h}}\}\\\Im\{\bar{\mathbf{h}}\}
\end{bmatrix}, \; \mathbf{z}_\mathrm{t} = \begin{bmatrix}
\Re\{\bar{\mathbf{z}}_\mathrm{t}\}\\\Im\{\bar{\mathbf{z}}_\mathrm{t}\}
\end{bmatrix}, \; \text{and}\\
\mathbf{P} &= \begin{bmatrix}
\Re \{\bar{\mathbf{P}}\} & -\Im \{\bar{\mathbf{P}}\}\\
\Im \{\bar{\mathbf{P}}\} & \Re \{\bar{\mathbf{P}}\}
\end{bmatrix}.
\end{align*}

\subsection{Bussang-based linear channel estimators}
We first revisit the Bussgang-based linear channel estimators including BMMSE and BWZF for low-resolution massive MIMO systems~\cite{li2017channel,Kolomvakis2020Quantized}.
The system model in~\eqref{eq_chan_est_real_vec_form} can be linearized by the Bussang decomposition as follows:
\begin{align}
    \mathbf{y}_\mathrm{t} & = \mathbf{V}_\mathrm{t}\mathbf{P}\mathbf{h} + \mathbf{V}_\mathrm{t}\mathbf{z}_\mathrm{t} + \mathbf{d}_\mathrm{t}\nonumber \\
    &=\mathbf{A}_\mathrm{t}\mathbf{h} + \mathbf{n}_\mathrm{t}
\end{align}
where the matrix $\mathbf{V}_\mathrm{t}$ is given as~\cite{Kolomvakis2020Quantized}
\begin{align*}
    \mathbf{V}_\mathrm{t} &= \frac{\Delta}{\sqrt{\pi}} \operatorname{diag}(\boldsymbol{\Sigma}_{\mathbf{r}_\mathrm{t}})^{-\frac{1}{2}} \times \\ 
    & \qquad \sum_{i=1}^{2^b - 1}\exp \bigg \{-\Delta^2 (i - 2^{b-1})^2\operatorname{diag}(\boldsymbol{\Sigma}_{\mathbf{r}_\mathrm{t}})^{-1} \bigg\}
\end{align*}
with $\boldsymbol{\Sigma}_{\mathbf{r}_\mathrm{t}} = \mathbf{P}\boldsymbol{\Sigma}_{\mathbf{h}}\mathbf{P}^T + \frac{N_0}{2}\mathbf{I}$ being the auto-correlation matrix of $\mathbf{r}_{\mathrm{t}}$. For the case of one-bit ADCs with $\Delta = \sqrt{2}$, the matrix $\mathbf{V}_\mathrm{t}$ reduces to a form as reported in~\cite[Eq. (10)]{li2017channel}.

%For the case of 1-bit ADCs, the covariance of $\mathbf{n}_\mathrm{t}$ is given in a closed form as
%\begin{equation}\label{Sigma-n}
%\begin{split}
%\boldsymbol{\Sigma}_{\mathbf{n}_\mathrm{t}} =& \frac{\Delta^2}{\pi}\Big[\operatorname{arcsin}\Big(\operatorname{diag}(\mathbf{\Sigma}_{\mathbf{r}_\mathrm{t}})^{-\frac{1}{2}}\mathbf{\Sigma}_{\mathbf{r}_\mathrm{t}}\operatorname{diag}(\mathbf{\Sigma}_{\mathbf{r}_\mathrm{t}})^{-\frac{1}{2}}\Big)-\\
%&\;\operatorname{diag}(\mathbf{\Sigma}_{\mathbf{r}_\mathrm{t}})^{-\frac{1}{2}}\mathbf{\Sigma}_{\mathbf{r}_\mathrm{t}}\operatorname{diag}(\mathbf{\Sigma}_{\mathbf{r}_\mathrm{t}})^{-\frac{1}{2}}+ \frac{N_0}{2}\operatorname{diag}(\mathbf{\Sigma}_{\mathbf{r}_\mathrm{t}})^{-1}\Big].
%\end{split}
%\end{equation}
%For $b$-bit ADCs with $b>1$, the covariance of $\mathbf{n}_\mathrm{t}$ can be approximated as $\boldsymbol{\Sigma}_{\mathbf{n}_\mathrm{t}} \approx (1-\eta_b)\eta_b\operatorname{diag}(\boldsymbol{\Sigma}_{\mathbf{r}_\mathrm{t}})$. Here $\boldsymbol{\Sigma}_{\mathbf{n}_\mathrm{t}}$ is often modeled as Gaussian noise as $\mathcal{N}(\mathbf{0},\boldsymbol{\Sigma}_{\mathbf{n}_\mathrm{t}})$.

The BMMSE channel estimator is given as~\cite{li2017channel,Kolomvakis2020Quantized}
\begin{align}
    \hat{\mathbf{h}}_{\mathtt{BMMSE}} = \boldsymbol{\Sigma}_{\mathbf{h}\mathbf{y}_\mathrm{t}} \boldsymbol{\Sigma}^{-1}_{\mathbf{y}_\mathrm{t}}\mathbf{y}_\mathrm{t} = \mathbf{A}^T_{\mathrm{t}}\boldsymbol{\Sigma}^{-1}_{\mathbf{y}_\mathrm{t}}\mathbf{y}_\mathrm{t}
\end{align}
where $\boldsymbol{\Sigma}_{\mathbf{h}\mathbf{y}_\mathrm{t}}$ is the cross-correlation matrix between $\mathbf{h}$ and $\mathbf{y}_\mathrm{t}$, and $\boldsymbol{\Sigma}_{\mathbf{y}_\mathrm{t}}$ is the auto-correlation matrix of $\mathbf{y}_{\mathrm{t}}$. For the case of one-bit ADCs, $\boldsymbol{\Sigma}_{\mathbf{y}_\mathrm{t}}$ is given as~\cite{li2017channel}
\begin{equation}
    \boldsymbol{\Sigma}_{\mathbf{y}_\mathrm{t}} = \frac{\Delta^2}{\pi}\operatorname{arcsin}\Big(\operatorname{diag}(\mathbf{\Sigma}_{\mathbf{r}_\mathrm{t}})^{-\frac{1}{2}}\mathbf{\Sigma}_{\mathbf{r}_\mathrm{t}}\operatorname{diag}(\mathbf{\Sigma}_{\mathbf{r}_\mathrm{t}})^{-\frac{1}{2}}\Big).
\end{equation}
For the case of few-bit ADCs, $\boldsymbol{\Sigma}_{\mathbf{y}_\mathrm{t}}$ is given as~\cite{Kolomvakis2020Quantized}
\begin{equation}
    \boldsymbol{\Sigma}_{\mathbf{y}_\mathrm{t}} = \mathbf{V}_{\mathrm{t}}\boldsymbol{\Sigma}_{\mathbf{r}_{\mathrm{t}}}\mathbf{V}^T_{\mathrm{t}} + \boldsymbol{\Sigma}_{\mathbf{d}_{\mathrm{t}}}.
\end{equation}
where $\boldsymbol{\Sigma}_{\mathbf{d}_{\mathrm{t}}}$ is the auto-correlation matrix of $\mathbf{d}_{\mathrm{t}}$ and can be approximated as $\boldsymbol{\Sigma}_{\mathbf{d}_{\mathrm{t}}}\approx \eta_b \operatorname{diag}(\boldsymbol{\Sigma}_{\mathbf{r}_{\mathrm{t}}})$. The distortion factor $\eta_b$ depending on the number of quantization bits $b$ is given in Table~\ref{table_standard_quantizer}.

A BWZF channel estimator was also proposed in~\cite{Kolomvakis2020Quantized} as follows:
\begin{equation}
    \hat{\mathbf{h}}_{\mathtt{BWZF}} = \big(\mathbf{A}^T_{\mathrm{t}}\operatorname{diag}(\mathbf{w})\mathbf{A}_{\mathrm{t}}\big)^{-1}\mathbf{A}^T_{\mathrm{t}}\operatorname{diag}(\mathbf{w})\mathbf{y}_{\mathrm{t}}
\end{equation}
where $\mathbf{w} = [w_1, w_2, \ldots, w_{2NT_\mathrm{t}}]^T$ with $$w_i = \frac{1}{\mathbb{E}[z^2_{\mathrm{t},i}] + \mathbb{E}[d^2_{\mathrm{t},i}|y_{\mathrm{t},i}]}, \; i = 1, \ldots, 2NT_\mathrm{t}.$$
Here, $y_{\mathrm{t},i}$, $z_{\mathrm{t},i}$, and $d_{\mathrm{t},i}$ are the $i$-th element of $\mathbf{y}_{\mathrm{t}}$, $\mathbf{z}_{\mathrm{t}}$, and $\mathbf{d}_{\mathrm{t}}$, respectively. The key idea of BZWF is that given an observed quantized signal vector $\mathbf{y}_{\mathrm{t}}$, the elements of $\mathbf{r}_{\mathrm{t}}$ have different variances. Exploiting this fact, the BWZF estimator sets the signals with lower variances to have higher weights.

\begin{table}[t!]
\caption{Optimum uniform quantizer for a Gaussian input as $\mc{C}(0, 1)$~\cite{max1960quantizing}.\label{table_standard_quantizer}}
\centering
\begin{tabular}{|l|c|c|c|c|}
\hline
\textbf{Resolution $b$}       & 1-bit & 2-bit & 3-bit & 4-bit \\ \hline
\textbf{Step size $\Delta_b$} & $\sqrt{8/\pi}$ & $0.996$ & $0.586$ & $0.335$ \\ \hline
\textbf{Distortion $\eta_b$}  & $1-2/\pi$ &  $0.1188$ & $0.0374$ &  $0.0115$ \\ \hline
\end{tabular}
\end{table}

\subsection{Proposed FBM-CENet}
\subsubsection{Maximum-likelihood channel estimation problem}
Let $\mathbf{P} = [\mathbf{p}_1,\mathbf{p}_2,\ldots,\mathbf{p}_{2NT_\mathrm{t}}]^T$, $\mathbf{y}_{\mathrm{t}} = [y_{\mathrm{t},1},\ldots,y_{\mathrm{t},2NT_{\mathrm{t}}}]^T$, and
$\mathbf{z}_{\mathrm{t}} = [z_{\mathrm{t},1},\ldots,z_{\mathrm{t},2NT_{\mathrm{t}}}]^T,
$
then we have 
\begin{equation}
y_{\mathrm{t},i} = \mathcal{Q}_b\left(\mathbf{p}_i^T\mathbf{h}+z_{\mathrm{t},i}\right), \quad i = 1,2,\ldots,2NT_\mathrm{t}.
\label{eq_chanEst_as_binary_classification}
\end{equation}
Let $s^\mathrm{up}_{\mathrm{t},i}=\sqrt{2\rho}(q^{\mathrm{up}}_{\mathrm{t},i}-\mathbf{p}_{i}^T\mathbf{h})$ and $s^\mathrm{low}_{\mathrm{t},i}=\sqrt{2\rho}(q^{\mathrm{low}}_{\mathrm{t},i}-\mathbf{p}_i^T\mathbf{h})$, where
\begin{align*}
    q^{\mathrm{up}}_{\mathrm{t},i} &= 
    \begin{cases}
    y_{\mathrm{t},i}+\frac{\Delta}{2} & \text{if}\; y_{\mathrm{t},i}<\tau_{2^b-1}\\
    \infty & \text{otherwise},
    \end{cases}\\
    q^{\mathrm{low}}_{\mathrm{t},i} &= 
    \begin{cases}
    y_{\mathrm{t},i}-\frac{\Delta}{2} & \text{if}\; y_{\mathrm{t},i}>\tau_1\\
    -\infty & \text{otherwise}.
    \end{cases}
\end{align*}
Hence, $q^{\mathrm{up}}_{\mathrm{t},i}$ and $q^{\mathrm{low}}_{\mathrm{t},i}$ are the upper and lower quantization thresholds of the bin to which $y_{\mathrm{t},i}$ belongs. 

The ML channel estimator is given as follows: 
\begin{eqnarray}
    \hat{\mathbf{h}}_{\texttt{ML}}  &=& \arg\max_{\mathbf{h}} \; f(\mathbf{y}_{\mathrm{t}}\,|\,\mathbf{h}) \nonumber \\
    &=& \arg\max_{\mathbf{h}} \; \sum_{i=1}^{2NT_\mathrm{t}}\log\left[\Phi \left(s^\mathrm{up}_{\mathrm{t},i}\right)-\Phi \left(s^\mathrm{low}_{\mathrm{t},i}\right)\right]. \label{eq_ML_chan_est_problem}
\end{eqnarray}

%\textcolor{red}{The ML estimation problem is a convex problem (there is a proof for this). Technically, it's possible to find the globally optimal solution for the ML estimator using numerical method. But the evaluation of the probit function can be a nuisance. That motivates us to use the logistic function and use deep unfolding to fold the numerical iterations.}

Let $\mathcal{P}_{\mathrm{t}}(\mathbf{h})$ be the objective function of~\eqref{eq_ML_chan_est_problem}. Since $\mathcal{P}_{\mathrm{t}}(\mathbf{h})$ is a concave function~\cite{pratt1981concavity}, the unconstrained optimization problem~\eqref{eq_ML_chan_est_problem} is convex, and therefore an iterative gradient ascent method can be used to solve~\eqref{eq_ML_chan_est_problem}. However, the gradient of $\mathcal{P}_{\mathrm{t}}(\mathbf{h})$, given by
\begin{equation}
    \nabla \mathcal{P}_{\mathrm{t}}(\mathbf{h}) = \sum_{i=1}^{2NT_\mathrm{t}} \frac{-\sqrt{2\rho}\mathbf{p}_i\big( \phi \left(s^\mathrm{up}_{\mathrm{t},i}\right)-\phi \left(s^\mathrm{low}_{\mathrm{t},i}\right) \big)}{\Phi \left(s^\mathrm{up}_{\mathrm{t},i}\right)-\Phi \left(s^\mathrm{low}_{\mathrm{t},i}\right)},
    \label{eq_original_channel_gradient}
\end{equation}
suffers from a vanishing issue, since the function $\Phi(\cdot)$ approaches zero or one very fast. Specifically, the iterative gradient descent method sequentially updates the estimated channel $\hat{\mathbf{h}}$. During the process of updating $\hat{\mathbf{h}}$, there exists an instance of $\hat{\mathbf{h}}$ that makes both $\Phi \left(s^\mathrm{up}_{\mathrm{t},i}\right)$ and $\Phi \left(s^\mathrm{low}_{\mathrm{t},i}\right)$ equal to zero or one. Thus, the denominator in~\eqref{eq_original_channel_gradient} can be zero for some $\hat{\mathbf{h}}$ causing the gradient vanishing issue. In addition, a lack of a closed-form expression for $\Phi(\cdot)$ complicates the evaluation in \eqref{eq_ML_chan_est_problem}. This observation motivates us to reformulate the ML channel estimation problem~\eqref{eq_ML_chan_est_problem} to address the vanishing issue as well as the complicated evaluation of the objective function in~\eqref{eq_ML_chan_est_problem}. We exploit a result in~\cite{bowling2009logistic}, which shows that the function $\Phi(t)$ can be accurately approximated by the Sigmoid function $\sigma(t) = 1/(1+e^{-t})$ as follows:
\begin{equation}
	\Phi(t) \approx \sigma(ct) = \frac{1}{1+e^{-ct}}
	\label{eq_approximate_Phi_as_Sigma}
\end{equation}
where $c = 1.702$ is a constant. It was shown in~\cite{bowling2009logistic} that $|\Phi(t)-\sigma(ct)|\leq 0.0095$, $\forall t\in \mathbb{R}$. The objective function $\mathcal{P}_{\mathrm{t}}(\mathbf{h})$ can be re-written as follows:
\begin{align}
    \mathcal{P}_{\mathrm{t}}(\mathbf{h}) &\approx  \tilde{\mathcal{P}}_{\mathrm{t}}(\mathbf{h}) = \sum_{i=1}^{2NT_\mathrm{t}}\log\left[\frac{1}{1+e^{-cs^\mathrm{up}_{\mathrm{t},i}}}-\frac{1}{1+e^{-cs^\mathrm{low}_{\mathrm{t},i}}}\right] \notag \\
    &= \sum_{i=1}^{2NT_\mathrm{t}}\Big[\log\Big(e^{-cs^\mathrm{low}_{\mathrm{t},i}}-e^{-cs^\mathrm{up}_{\mathrm{t},i}}\Big) - \log\left(1+e^{-cs^\mathrm{up}_{\mathrm{t},i}}\right) \notag\\ & \qquad \quad \;\;\;  -\log\left(1+e^{-cs^\mathrm{low}_{\mathrm{t},i}}\right)\Big].
\end{align}
Thus, a reformulated ML channel estimation problem is obtained as follows:
\begin{equation}
\hat{\mathbf{h}} = \arg \max_{\mathbf{h}}\; \tilde{\mathcal{P}}_{\mathrm{t}}(\mathbf{h}).
\label{eq_reformulated_MAP_chan_est_problem}
\end{equation}

The gradient of $\tilde{\mathcal{P}}_{\mathrm{t}}(\mathbf{h})$ is 
\begin{align}
    \nabla\tilde{\mathcal{P}}_{\mathrm{t}}(\mathbf{h})&=\sum_{i=1}^{2NT_\mathrm{t}}c\sqrt{2\rho}\,\mathbf{p}_i\left(1-\frac{1}{1+e^{cs^\mathrm{up}_{\mathrm{t},i}}}-\frac{1}{1+e^{cs^\mathrm{low}_{\mathrm{t},i}}}\right) \nonumber\\
    &= c\sqrt{2\rho}\,\mathbf{P}^T\Big[\mathbf{1}-\sigma\left(c\sqrt{2\rho}\left(\mathbf{P}\mathbf{h}-\mathbf{q}^{\mathrm{up}}_{\mathrm{t}}\right)\right)-\notag\\
    &\hspace{2.5cm}\sigma\left(c\sqrt{2\rho}\left(\mathbf{P}\mathbf{h}-\mathbf{q}^{\mathrm{low}}_{\mathrm{t}}\right)\right)\Big]
    \label{eq_reformulated_channel_gradient}
\end{align}
where $\mathbf{q}^{\mathrm{up}}_\mathrm{t} = [q_{\mathrm{t},1}^{\mathrm{up}},\ldots,q_{\mathrm{t},2NT_\mathrm{t}}^{\mathrm{up}}]^T$ and $\mathbf{q}^{\mathrm{low}}_\mathrm{t} = [q_{\mathrm{t},1}^{\mathrm{low}},\ldots,q_{\mathrm{t},2NT_\mathrm{t}}^{\mathrm{low}}]^T$. It can be seen that the gradient of $\tilde{\mathcal{P}}_{\mathrm{t}}(\mathbf{h})$ in~\eqref{eq_reformulated_channel_gradient} does not suffer from the divided-by-zero issue as in the gradient of $\mathcal{P}_{\mathrm{t}}(\mathbf{h})$. Thus, an iterative gradient decent method for solving~\eqref{eq_reformulated_MAP_chan_est_problem} can be written as
\begin{equation}
    \mathbf{h}^{(\ell)} = \mathbf{h}^{(\ell-1)} + \alpha^{(\ell)}_\mathrm{t} \nabla\tilde{\mathcal{P}}_{\mathrm{t}}\big(\mathbf{h}^{(\ell-1)}\big)
    \label{eq_iterative_channel_method}
\end{equation}
where $\ell$ is the iteration index and $\alpha^{(\ell)}_\mathrm{t}$ is the step size.
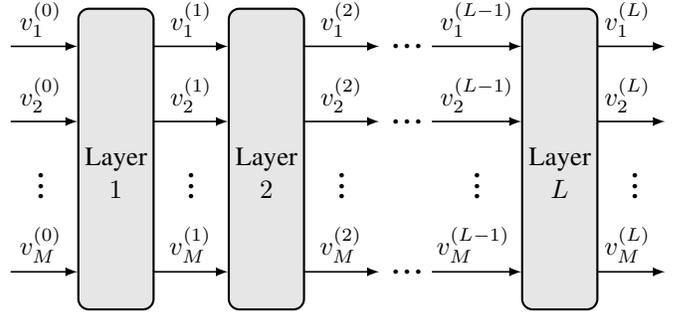
\begin{figure}[t!]
	\centering
	\begin{tikzpicture}
	\node [above] at (0.4,10) {$v^{(0)}_1$};
	\draw [semithick,->] (0,10) to (0.89,10);
	
	\node [above] at (0.4,9) {$v^{(0)}_2$};
	\draw [semithick,->] (0,9) to (0.89,9);
	
	\node [above] at (0.4,7) {$v^{(0)}_{M}$};
	\draw [semithick,->] (0,7) to (0.89,7);
	
	\draw [fill=LayerColor, thick, rounded corners] (0.9,6.5) rectangle (1.9,10.5);
	\node at (1.4,8.5) {Layer};
	\node at (1.4,8.1) {$1$};
	
	\node [above] at (2.4,10) {$v^{(1)}_1$};
	\draw [semithick,->] (1.9,10) to (2.89,10);
	
	\node [above] at (2.4,9) {$v^{(1)}_2$};
	\draw [semithick,->] (1.9,9) to (2.89,9);
	
	\node [above] at (2.39,7) {$v^{(1)}_{M}$};
	\draw [semithick,->] (1.9,7) to (2.89,7);
	
	\draw [fill=LayerColor, thick, rounded corners] (2.9,6.5) rectangle (3.9,10.5);
	\node at (3.4,8.5) {Layer};
	\node at (3.4,8.1) {$2$};
	
	\node [above] at (4.4,10) {$v^{(2)}_1$};
	\draw [semithick,->] (3.9,10) to (4.9,10);
	
	\node [above] at (4.4,9) {$v^{(2)}_2$};
	\draw [semithick,->] (3.9,9) to (4.9,9);
	
	\node [above] at (4.39,7) {$v^{(2)}_{M}$};
	\draw [semithick,->] (3.9,7) to (4.9,7);
	
	\node at (5.3,10) {\textbf{\ldots}};
	\node at (5.3,9) {\textbf{\ldots}};
	\node at (5.3,7) {\textbf{\ldots}};
	
	\node [above] at (6.2,10) {$v^{(L-1)}_1$};
	\draw [semithick,->] (5.6,10) to (6.79,10);
	
	\node [above] at (6.2,9) {$v^{(L-1)}_2$};
	\draw [semithick,->] (5.6,9) to (6.79,9);
	
	\node [above] at (6.15,7) {$v^{(L-1)}_{M}$};
	\draw [semithick,->] (5.6,7) to (6.79,7);
	
	\draw [fill=LayerColor, thick, rounded corners] (6.8,6.5) rectangle (7.8,10.5);
	\node at (7.3,8.5) {Layer};
	\node at (7.3,8.1) {$L$};
	
	\node [above] at (8.2,10) {$v^{(L)}_1$};
	\draw [semithick,->] (7.8,10) to (8.7,10);
	\node [above] at (8.2,9) {$v^{(L)}_2$};
	\draw [semithick,->] (7.8,9) to (8.7,9);
	\node [above] at (8.2,7) {$v^{(L)}_{M}$};
	\draw [semithick,->] (7.8,7) to (8.7,7);
	
	\node at (0.4,8.25) {\textbf{\vdots}};
	\node at (2.4,8.25) {\textbf{\vdots}};
	\node at (4.4,8.25) {\textbf{\vdots}};
	\node at (6.2,8.25) {\textbf{\vdots}};
	\node at (8.3,8.25) {\textbf{\vdots}};
	\end{tikzpicture}
	\caption{Overall structure of the proposed FBM-CENet, FBM-DetNet, and B-DetNet. For FBM-CENet, $v$ plays the role of $h$ and $M = 2NK$. For FBM-DetNet and B-DetNet, $v$ plays the role of $x$ and $M = 2K$.}
	\label{fig_overall_network_structure}
\end{figure}

\subsubsection{Network structure of the proposed FBM-CENet}
We employ the deep unfolding technique~\cite{Hershey-Unfolding-2014} to unfold each iteration in~\eqref{eq_iterative_channel_method} as a layer of a deep neural network. The overall structure of the proposed FBM-CENet estimator is illustrated in Fig.~\ref{fig_overall_network_structure}, where there are $L$ layers and each layer takes a vector of $2NK$ elements as the input and generates an output vector of the same size. 

The specific structure for each layer $\ell$ of the proposed FBM-CENet is illustrated in Fig.~\ref{fig_proposed_chanest_layer_structure}. The proposed layer structure is special and unique due to the use of the approximation  in~\eqref{eq_approximate_Phi_as_Sigma} and the structure of the reformulated gradient in~\eqref{eq_reformulated_channel_gradient}. Specifically, each layer of the proposed FBM-CENet consists of two weight matrices and two bias vectors where the pilot matrix $\mathbf{P}$ plays the role of the weight matrices and the received signals $\mathbf{q}^{\mathrm{up}}_\mathrm{t}$ and $\mathbf{q}^{\mathrm{low}}_\mathrm{t}$ play the role of the bias vectors. By contrast, each layer $\ell$ of a conventional DNN-based channel estimator as illustrated in Fig.~\ref{fig_conventional_DNN_layer_structure} contains one weight matrix $\mathbf{W}_\ell$ and one bias vector $\mathbf{b}_\ell$. Such a conventional DNN structure has been employed in several existing works, e.g.,~\cite{DuyNguyen2020Neural,Gao2019Deep,Zicheng2021Deep}. An interesting feature of the proposed network structure is the Sigmoid activation function $\sigma(\cdot)$, which is not arbitrary but results from the use of the approximation in~\eqref{eq_approximate_Phi_as_Sigma}. This is unlike the conventional DNN structure where the activation functions $\{f_\ell(\cdot)\}$ are often chosen heuristically by experiments. 

\subsubsection{Trainable parameters}
For a given pilot matrix $\mathbf{P}$, the trainable parameters in the proposed FBM-CENet are the step sizes $\{\alpha^{(\ell)}_\mathrm{t}\}$ and a scaling parameter $\beta_\mathrm{t}$ inside the Sigmoid function. Note that the coefficient $c\sqrt{2\rho}$ is omitted in the proposed network structure since it is a constant through all the layers of the network. The trainable parameters $\{\alpha^{(\ell)}_\mathrm{t}\}$ and $\beta_\mathrm{t}$ take over the role of this coefficient.

\begin{figure}[t!]
	\centering
	\begin{subfigure}[t]{0.48\textwidth}
	\centering
	\begin{tikzpicture}
	\node at (1.2,5) {$\mathbf{y}_{\mathrm{t}}$};
	\draw [semithick,->] (1.4,5) to (1.9,5);
	\draw [semithick] (2.1,5) circle [radius=0.2];
	\node at (2.1,5) {$\times$};
	\node at (2.15,5.88) {$\mathbf{W}_1$};
	\draw [semithick,->] (2.1,5.7) to (2.1,5.2);
	\draw [semithick,->] (2.3,5) to (3,5);
	\node at (2.68,5.22) {$\boldsymbol{\theta}_{1}$};
	\draw [semithick] (3,4.7) rectangle (5,5.3);
	\node at (4,5) {$f_1\big(\boldsymbol{\theta}_{1}-\mathbf{b}_1\big)$};
    % another layer
	\draw [semithick,->] (5,5) to (5.7,5);
	\draw [semithick] (5.9,5) circle [radius=0.2];
	\node at (5.9,5) {$\times$};
	\node at (5.95,5.88) {$\mathbf{W}_2$};
	\draw [semithick,->] (5.9,5.7) to (5.9,5.2);
	\draw [semithick,->] (6.1,5) to (6.8,5);
	\node at (6.48,5.22) {$\boldsymbol{\theta}_{2}$};
	\draw [semithick] (6.8,4.7) rectangle (8.8,5.3);
	\node at (7.8,5) {$f_2\big(\boldsymbol{\theta}_{2}-\mathbf{b}_2\big)$};
	\draw[semithick,->] (8.8,5) to (9.2,5) to (9.2, 5) to (9.2, 3.4) to (8.6, 3.4);
	\node at (8.3, 3.4) {\textbf{$\ldots$}};
	\draw[semithick,->] (8,3.4) to (7.4, 3.4);
	\draw [semithick] (7.2,3.4) circle [radius=0.2];
	\node at (7.2,3.4) {$\times$};
	\node at (7.3,4.25) {$\mathbf{W}_{L}$};
	\draw[semithick,->] (7.2,4.1) to (7.2,3.6);
	\draw[semithick,->] (7,3.4) to (6.1,3.4);
	\node at (6.6,3.6) {$\boldsymbol{\theta}_{L}$};
	\draw [semithick] (3.9,3.1) rectangle (6.1,3.7);
	\node at (5,3.4) {$f_L\big(\boldsymbol{\theta}_{L}-\mathbf{b}_{L}\big)$};
	\draw[semithick,->] (3.9,3.4) to (3.2,3.4);
	\node at (3,3.5) {$\hat{\mathbf{h}}$};
	\end{tikzpicture}
	\caption{Conventional channel estimation DNN structure. Each layer $\ell$ contains a trainable weight matrix $\mathbf{W}_{\ell}$, a trainable bias vector $\mathbf{b}_{\ell}$, and an activation function $f_{\ell}(\cdot)$.}
	\label{fig_conventional_DNN_layer_structure}
	\end{subfigure}
	\begin{subfigure}[t]{0.48\textwidth}
	\centering
	\begin{tikzpicture}
	\node at (1.3,5.22) {$\mathbf{h}^{(\ell-1)}$};
	\draw [semithick,->] (0.8,5) to (1.9,5);
	\draw [semithick] (2.1,5) circle [radius=0.2];
	\node at (2.1,5) {$\times$};
	\node at (2.15,5.9) {$\mathbf{P}$};
	\draw [semithick,->] (2.1,5.7) to (2.1,5.2);
	\draw [semithick,-] (2.3,5) to (3,5);
	\node at (2.65,5.27) {$\mathbf{u}_{\mathrm{t}}^{(\ell)}$};
	\draw [semithick,<->] (3.35,5.7) to (3,5.7) to (3,4.3) to (3.35,4.3);
	\draw [semithick] (3.35,5.4) rectangle (6.16,6);
	\node at (4.77,5.7) {$\sigma\big(\beta_\mathrm{t}(\mathbf{u}_{\mathrm{t}}^{(\ell)}-\mathbf{q}^{\mathrm{up}}_\mathrm{t})\big)$};
	\draw [semithick] (3.35,4) rectangle (6.16,4.6);
	\node at (4.77,4.3) {$\sigma\big(\beta_\mathrm{t}(\mathbf{u}_{\mathrm{t}}^{(\ell)}-\mathbf{q}^{\mathrm{low}}_\mathrm{t})\big)$};
	\draw [semithick,->] (6.17,5.7) to (6.4,5.7) to (6.4,5.2);
	\draw [semithick,->] (6.16,4.3) to (6.4,4.3) to (6.4,4.8);
	\draw [semithick] (6.4,5) circle [radius=0.2];
	\draw [semithick,->] (5.7,5) to (6.2,5);
	\node at (5.6,5) {$\mathbf{1}$};
	\node at (6.1,5.18) {$+$};
	\node at (6.6,5.3) {$-$};
	\node at (6.6,4.7) {$-$};
	\draw [semithick,->] (6.6,5) to (7.1,5);
	\draw [semithick] (7.3,5) circle [radius=0.2];
	\node at (7.3,5) {$\times$};
	\draw [semithick,->] (7.3,5.7) to (7.3,5.2);
	\node at (7.42,5.9) {$\mathbf{P}^T$};
	\draw [semithick,->] (7.3,4.3) to (7.3,4.8);
	\node at (7.45,4.2) {$\alpha^{(\ell)}_\mathrm{t}$};
	\draw [semithick,->] (7.5,5) to (8,5);
	\draw [semithick] (8.2,5) circle [radius=0.2];
	\draw [semithick,->] (8.2,5.7) to (8.2,5.2);
	\node at (8.55,5.9) {$\mathbf{h}^{(\ell-1)}$};
	\node at (8.2,5) {$+$};
	\draw [semithick,->] (8.4,5) to (9,5);
	\node at (9,5.25) {$\mathbf{h}^{(\ell)}$};
	\end{tikzpicture}
	\caption{Specific structure of layer $\ell$ of the proposed FBM-CENet.}
	\label{fig_proposed_chanest_layer_structure}
	\end{subfigure}
	\caption{Conventional versus proposed DNN structure for channel estimation.}
	\label{fig_conventional_vs_proposed_DNN_layer_structure}
\end{figure}
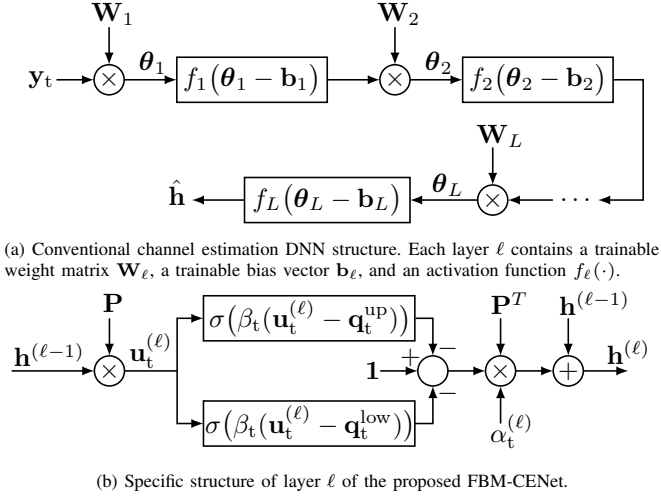

It is important to note that the pilot matrix $\mathbf{P}$ directly plays the role of the weight matrices. Therefore, when the pilot matrix $\mathbf{P}$ is not given, it can be treated as a trainable parameter. In this case, training the proposed FBM-CENet is equivalent to \textit{jointly optimizing both the channel estimator at the base station and the pilot signal transmitted from the users}. This is a significant advantage of the proposed network structure since the conventional DNN-based channel estimator is often trained or optimized for a given pilot matrix. In other words, conventional DNN structures do not convey information about the optimal pilot signal. We note that a recent work in~\cite{DuyNguyen2020Neural} also jointly optimized the pilot signal and the channel estimator for massive MIMO systems with low-resolution ADCs. However, the channel estimator in~\cite{DuyNguyen2020Neural} simply employs the conventional DNN structure as illustrated in Fig.~\ref{fig_conventional_DNN_layer_structure}. We will later show that the proposed FBM-CENet estimator significantly outperforms the method in~\cite{DuyNguyen2020Neural}.

\subsubsection{Training strategy}
Here we present the strategy for straining the proposed FBM-CENet estimator. Let $\hat{\mathbf{h}}$ denote the channel estimate, which is set to be the output of the last layer of the proposed FBM-CENet, i.e., $\hat{\mathbf{h}}=\mathbf{h}^{(L)}$. The cost function to be minimized is $\|\hat{\mathbf{h}} - \mathbf{h}\|^2$. 

In case the pilot matrix $\mathbf{P}$ is given, a training sample for the proposed FBM-CENet contains the given matrix $\mathbf{P}$, a channel vector realization $\mathbf{h}$ and a noise vector $\mathbf{z}$, which can be obtained by random generating. When the pilot matrix $\mathbf{P}$ is not given and it is trainable, a training sample only contains a channel vector realization $\mathbf{h}$ and a noise vector $\mathbf{z}$.

\begin{figure}[t!]
    \centering
    \begin{subfigure}[t]{0.24\textwidth}
        \centering
        \includegraphics[width=\linewidth]{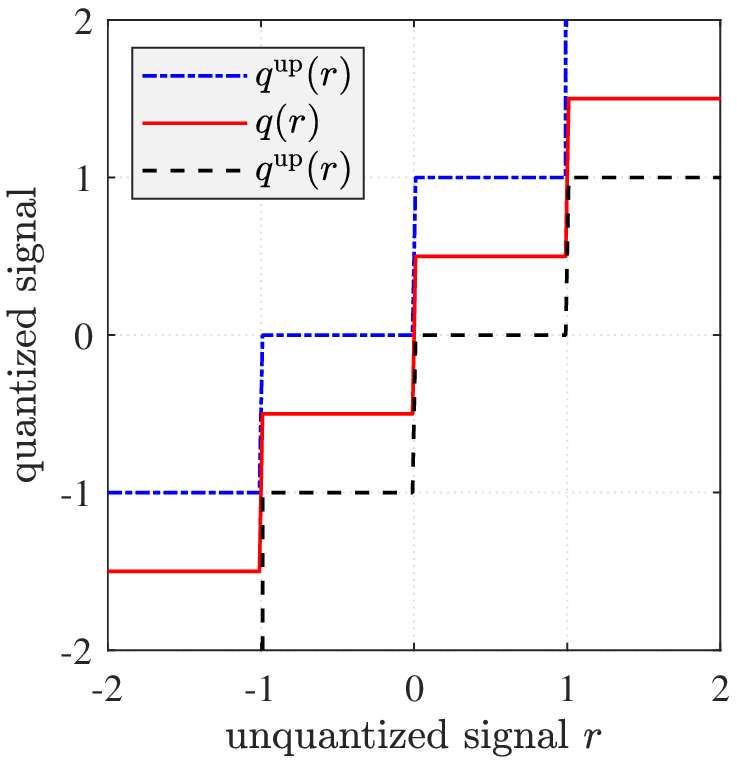}
        \caption{$c_1 = 0.01$.}
    \end{subfigure}~
    \begin{subfigure}[t]{0.24\textwidth}
        \centering
        \includegraphics[width=\linewidth]{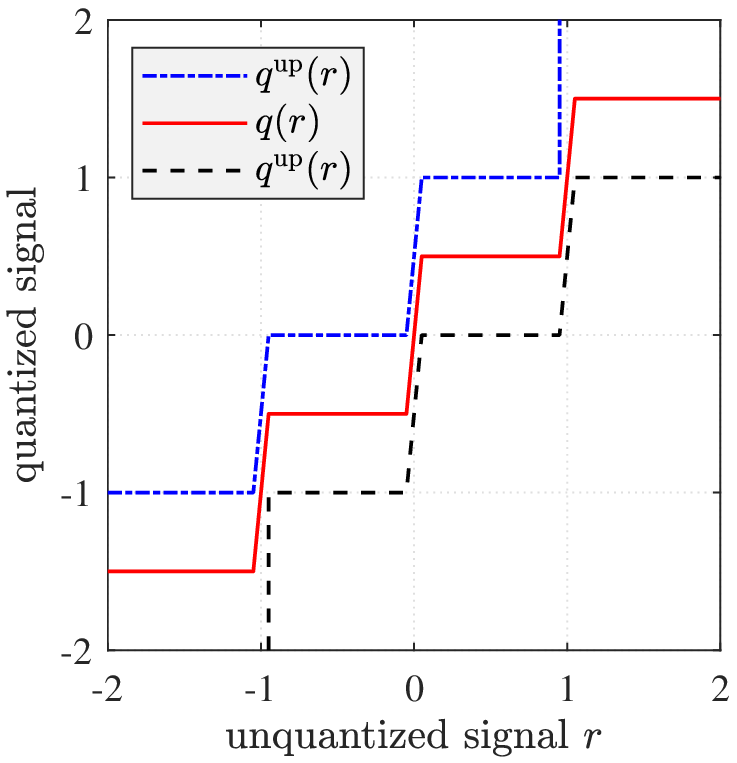}
        \caption{$c_1 = 0.05$}
    \end{subfigure}
    \caption{Two-bit soft quantizer with $\Delta = 1$.}
    \label{fig_soft_quantizer}
\end{figure}

It is important to note that the received signals $\mathbf{q}_\mathrm{t}^{\mathrm{up}}$ and $\mathbf{q}_\mathrm{t}^{\mathrm{low}}$ depend on the pilot matrix $\mathbf{P}$. Therefore, in case the pilot matrix $\mathbf{P}$ is trainable, gradient back-propagation during the training process should also go through $\mathbf{q}_\mathrm{t}^{\mathrm{up}}$ and $\mathbf{q}_\mathrm{t}^{\mathrm{low}}$. However, the low-resolution ADCs are discontinuous functions, which make gradient back-propagation through $\mathbf{q}_\mathrm{t}^{\mathrm{up}}$ and $\mathbf{q}_\mathrm{t}^{\mathrm{low}}$ infeasible. To overcome this issue, we employ a soft quantizer model based on the Rectified Linear Unit (ReLU) activation function $f_{\mathrm{relu}}(r) = \max(0,r)$ for the training process as follows: 
\begin{align}
    q^{\mathrm{up}}(r) & = q(r) + \frac{\Delta}{2} +  c_2\big[f_{\mathrm{relu}}(r - B\Delta + c_1) - \notag \\
    & \qquad f_{\mathrm{relu}}(r - B\Delta - c_1)\big] \label{eq_soft_qup_based_on_relu}\\
    q^{\mathrm{low}}(r) & = q(r) - \frac{\Delta}{2} -  c_2\big[f_{\mathrm{relu}}(-r - B\Delta + c_1) - \notag \\
    & \qquad f_{\mathrm{relu}}(-r - B\Delta - c_1)\big] 
    \label{eq_soft_qlow_based_on_relu}
\end{align}
where $B = 2^{b-1}-1$, $c_1$ and $c_2$ are positive constants, and 
\begin{align}
    q(r) & = -(2^b - 1)\frac{\Delta}{2} + 
    \frac{\Delta}{2c_1}  \sum_{i=-B}^{B} \big[f_{\mathrm{relu}}(r + i\Delta + c_1) -  \notag \\
    & \qquad f_{\mathrm{relu}}(r + i\Delta - c_1)\big].
    \label{eq_soft_quantizer_based_on_ReLU}
\end{align}
This soft quantization model is based on the ReLU function, which is continuous and therefore back-propagation is feasible. The effect of $c_1$ is illustrated in Fig.~\ref{fig_soft_quantizer}. It can be seen that the smaller $c_1$ is, the sharper the soft quantizer is, or in other words, the closer the soft quantizer is to the hard (real) quantizer. The constant $c_2$ accounts for the two thresholds $\tau_0 = -\infty$ and $\tau_{2^b} = \infty$, and hence it should be a large number.

It should be noted that the constants $\{c_1, c_2\}$ should not be treated as trainable parameters because we need the soft quantizer to be close to the hard quantizer. If these constants are treated as trainable parameters, the training process may produce a soft quantizer that significantly deviates from the hard quanizer, which is in fact the model in the real systems.
%% ------------------- RESERVED ---------------
%  please do not delete the following equations
% ---------------------------------------------
%\begin{align}
%    q(r) &= \frac{\Delta}{2}\big[\sigma(c_1r) - \sigma(-c_1r)\big] + \notag \\
%    & \quad \; \; \Delta\sum_{i=1}^{B} \sigma(c_1(r-i\Delta)) - \sigma(c_1(-r-i\Delta)) \\
%    q^{\mathrm{up}}(r) & = q(r) + \frac{\Delta}{2} + c_2\sigma(c_1(r-B\Delta))\\
%    q^{\mathrm{low}}(r) & = q(r) - \frac{\Delta}{2} - c_2\sigma(c_1(-r-B\Delta)).
%\end{align}
%% ------------------------------------------

\section{Data Detection in Few-Bit MIMO Systems}
\label{sec_data_detection}
In this section, we propose two DNN-based detectors, namely B-DetNet and FBM-DetNet, for massive MIMO systems with low-resolution ADCs. For convenience in later derivations, we convert~\eqref{eq_analog_complex_received_signal} and~\eqref{eq_quantized_complex_received_signal} into the real domain as follows:
\begin{equation}
\mathbf{y} = \mathcal{Q}_b\left(\mathbf{H}\mathbf{x} + \mathbf{z}\right),
\label{eq_quantized_real_received_signal}
\end{equation}
where
\begin{align*}
\mathbf{y} &= \begin{bmatrix}
\Re \{\bar{\mathbf{y}}\} \\ \Im \{\bar{\mathbf{y}}\}
\end{bmatrix},\ \mathbf{x} = \begin{bmatrix}
\Re \{\bar{\mathbf{x}}\} \\ \Im \{\bar{\mathbf{x}}\}
\end{bmatrix}, \
\mathbf{z} = \begin{bmatrix}
\Re \{\bar{\mathbf{z}}\} \\ \Im \{\bar{\mathbf{z}}\}
\end{bmatrix}, \ \text{and}\\
\mathbf{H} &= \begin{bmatrix}
\Re \{\bar{\mathbf{H}}\} & -\Im \{\bar{\mathbf{H}}\}\\
\Im \{\bar{\mathbf{H}}\} & \Re \{\bar{\mathbf{H}}\}
\end{bmatrix}.
\end{align*}
Note that $\mathbf{y}\in \mathbb{R}^{2N}$, $\mathbf{x}\in \mathbb{R}^{2K}$, $\mathbf{z}\in \mathbb{R}^{2N}$, and $\mathbf{H}\in \mathbb{R}^{2N\times2K}$. We also denote $\mathbf{y} = [y_1, \ldots, y_{2N}]^T$ and $\mathbf{H} = [\mathbf{h}_1, \ldots, \mathbf{h}_{2N}]^T$.

\begin{figure}[t!]
    \centering
    \begin{subfigure}[t]{0.24\textwidth}
        \centering
        \includegraphics[width=\linewidth]{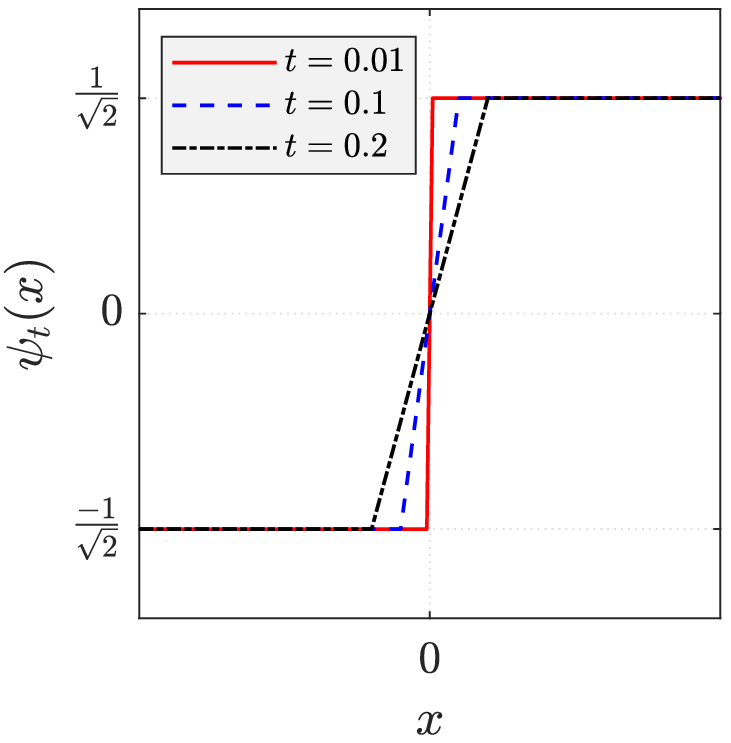}
        \caption{QPSK signaling.}
        \label{fig_soft_QPSK_projector}
    \end{subfigure}~
    \begin{subfigure}[t]{0.24\textwidth}
        \centering
        \includegraphics[width=\linewidth]{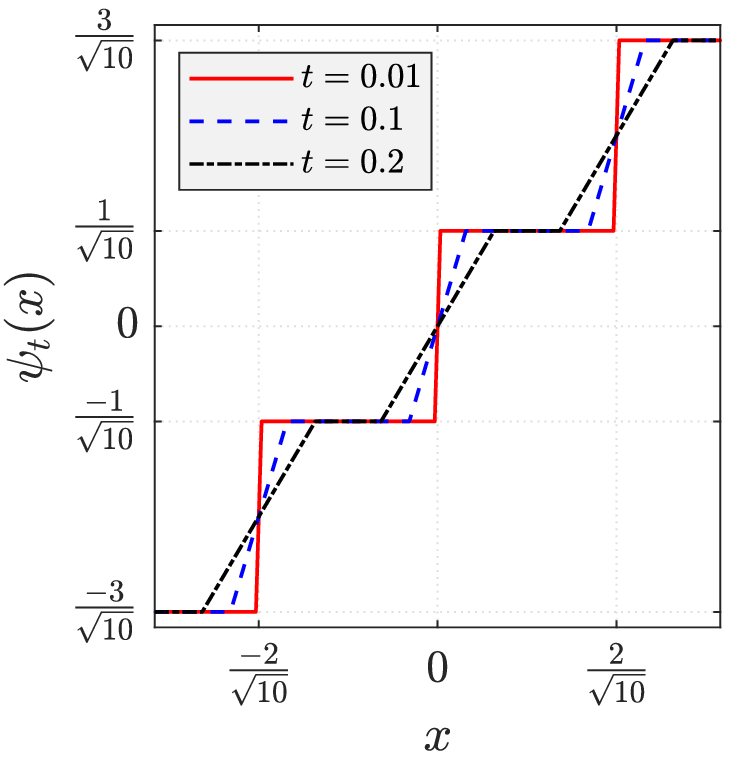}
        \caption{$16$QAM signaling.}
        \label{fig_soft_16QAM_projector}
    \end{subfigure}
    \caption{Projector function $\psi_t(\cdot)$ with different values of $t$.}
    \label{fig_soft_projector}
\end{figure}
\begin{figure*}[t!]
	\centering
	\begin{tikzpicture}
	\draw [fill=InOutColor, dashed, semithick, rounded corners] (2.2,6.5) rectangle (3.7,10.5);
	\draw [semithick,->] (2.95,5.9) to (2.95,6.5);
	\node at (2.95,5.7) {Input};
	
	\draw [fill=InOutColor, dashed, semithick, rounded corners] (15.8,6.3) rectangle (17.1,10.7);
	\draw [semithick,->] (16.5,5.6) to (16.5,6.3);
	\node at (16.5,5.4) {Output};
	
	\node at (5.3,11) {\small {weight matrix}};
	\node at (5.3,10.7) {\small {$\mathbf{A}$}};
	\node at (10.15,11.35) {\small {weight matrix}};
	\node at (10.15,11) {\small {$\mathbf{A}^T\boldsymbol{\Sigma}_{\mathbf{n}}^{-1}$}};
	
	\node [left] at (3.5,10) {$x^{(\ell-1)}_{1}$};
	\draw [semithick] (3.5,10) to (4.2,10);
	\draw [fill] (4.2,10) circle [radius=0.08];
	
	\draw [semithick, ->] (4.2,10) to (6.69,11);
	\draw [semithick, ->] (4.2,9) to (6.7,10.88);
	\draw [semithick, ->] (4.2,7) to (6.78,10.75);
	
	\draw [semithick] (7.05,11) circle [radius=0.35];
	\node at (7.05,11) {$\sum$};
	\draw [semithick,->] (7.4,11) to (8,11);
	\draw [semithick] (8.2,11) circle [radius=0.2];
	\draw [semithick,->] (8.2,10.3) to (8.2,10.79);
	\node at (8.2,10.1) {{\small $y_1$}};
	\node at (7.88,11.17) {{\small $-$}};
	\node at (8.4,10.65) {{\small $+$}};
	\draw [semithick] (8.4,11) to (8.7,11);
	\draw [fill] (8.7,11) circle [radius=0.08];
	
	\draw [semithick,->] (8.7,11) to (11.2,10.4);
	\draw [semithick,->] (8.7,11) to (11.2,9.1);
	\draw [semithick,->] (8.7,11) to (11.24,6.95);
	
	\draw [semithick] (11.5,10.2) circle [radius=0.35];
	\node at (11.5,10.2) {$\sum$};
	\draw [semithick,->] (11.85,10.2) to (12.49,10.2);
	\draw [semithick] (12.7,10.2) circle [radius=0.2];
	\node at (12.7,10.2) {$\times$};
	\draw [semithick,->] (12.7,9.6) to (12.7,9.99);
	\node at (12.9,9.46) {$\alpha^{(\ell)}$};
	\draw [semithick,->] (12.9,10.2) to (13.49,10.2);
	\draw [semithick] (13.7,10.2) circle [radius=0.2];
	\node at (13.7,10.2) {$+$};
	\draw [semithick,->] (13.7,9.6) to (13.7,9.99);
	\node at (14.05,9.45) {$x^{(\ell-1)}_{1}$};
	\draw [semithick,->] (13.9,10.2) to (14.6,10.2);
	\draw [semithick] (14.6,9.9) rectangle (15.6,10.5);
	\node at (15.1,10.2) {$\psi_{t_\ell}(\cdot)$};
	\draw [semithick,->] (15.6,10.2) to (16.2,10.2);
	\node at (16.5,10.25) {$x^{(\ell)}_{1}$};
	
	\node [left] at (3.5,9) {$x^{(\ell-1)}_{2}$};
	\draw [semithick] (3.5,9) to (4.2,9);
	\draw [fill] (4.2,9) circle [radius=0.08];
	
	\draw [semithick, ->] (4.2,10) to (6.75,9.2);
	\draw [semithick, ->] (4.2,9) to (6.7,9);
	\draw [semithick, ->] (4.2,7) to (6.75,8.8);
	
	\draw [semithick] (7.05,9) circle [radius=0.35];
	\node at (7.05,9) {$\sum$};
	\draw [semithick,->] (7.4,9) to (8,9);
	\draw [semithick] (8.2,9) circle [radius=0.2];
	\draw [semithick,->] (8.2,8.3) to (8.2,8.79);
	\node at (8.2,8.1) {{\small $y_2$}};
	
	\node at (7.88,9.17) {{\small $-$}};
	\node at (8.4,8.65) {{\small $+$}};
	\draw [semithick] (8.4,9) to (8.7,9);
	\draw [fill] (8.7,9) circle [radius=0.08];
	
	\draw [semithick,->] (8.7,9) to (11.14,10.2);
	\draw [semithick,->] (8.7,9) to (11.15,8.9);
	\draw [semithick,->] (8.7,9) to (11.15,6.77);
	
	\draw [semithick] (11.5,8.9) circle [radius=0.35];
	\node at (11.5,8.9) {$\sum$};
	\draw [semithick,->] (11.85,8.9) to (12.49,8.9);
	\draw [semithick] (12.7,8.9) circle [radius=0.2];
	\node at (12.7,8.9) {$\times$};
	\draw [semithick,->] (12.7,8.3) to (12.7,8.69);
	\node at (12.9,8.16) {$\alpha^{(\ell)}$};
	\draw [semithick,->] (12.9,8.9) to (13.49,8.9);
	\draw [semithick] (13.7,8.9) circle [radius=0.2];
	\node at (13.7,8.9) {$+$};
	\draw [semithick,->] (13.7,8.3) to (13.7,8.69);
	\node at (14.05,8.15) {$x^{(\ell-1)}_{2}$};
	\draw [semithick,->] (13.9,8.9) to (14.6,8.9);
	\draw [semithick] (14.6,8.6) rectangle (15.6,9.2);
	\node at (15.1,8.9) {$\psi_{t_\ell}(\cdot)$};
	\draw [semithick,->] (15.6,8.9) to (16.2,8.9);
	\node at (16.5,8.95) {$x^{(\ell)}_{2}$};
	
	\node [left] at (3.5,7) {$x^{(\ell-1)}_{2K}$};
	\draw [semithick] (3.5,7) to (4.2,7);
	\draw [fill] (4.2,7) circle [radius=0.08];
	
	\draw [semithick, ->] (4.2,10) to (6.79,6.25);
	\draw [semithick, ->] (4.2,9) to (6.7,6.1);
	\draw [semithick, ->] (4.2,7) to (6.7,5.92);
	
	\draw [semithick] (7.05,6) circle [radius=0.35];
	\node at (7.05,6) {$\sum$};
	\draw [semithick,->] (7.4,6) to (8,6);
	\draw [semithick] (8.2,6) circle [radius=0.2];
	\draw [semithick,->] (8.2,5.3) to (8.2,5.79);
	\node at (8.2,5.1) {{\small $y_{2N}$}};
	\node at (7.88,6.17) {{\small $-$}};
	\node at (8.4,5.65) {{\small $+$}};
	\draw [semithick] (8.4,6) to (8.7,6);
	\draw [fill] (8.7,6) circle [radius=0.08];
	
	\draw [semithick,->] (8.7,6) to (11.2,10);
	\draw [semithick,->] (8.7,6) to (11.2,8.7);
	\draw [semithick,->] (8.7,6) to (11.2,6.5);
	
	\draw [semithick] (11.5,6.7) circle [radius=0.35];
	\node at (11.5,6.7) {$\sum$};
	\draw [semithick,->] (11.85,6.7) to (12.49,6.7);
	\draw [semithick] (12.7,6.7) circle [radius=0.2];
	\node at (12.7,6.7) {$\times$};
	\draw [semithick,->] (12.7,6.1) to (12.7,6.49);
	\node at (12.9,5.96) {$\alpha^{(\ell)}$};
	\draw [semithick,->] (12.9,6.7) to (13.49,6.7);
	\draw [semithick] (13.7,6.7) circle [radius=0.2];
	\node at (13.7,6.7) {$+$};
	\draw [semithick,->] (13.7,6.1) to (13.7,6.49);
	\node at (14.05,5.95) {$x^{(\ell-1)}_{2K}$};
	\draw [semithick,->] (13.9,6.7) to (14.6,6.7);
	\draw [semithick] (14.6,6.4) rectangle (15.6,7.0);
	\node at (15.1,6.7) {$\psi_{t_\ell}(\cdot)$};
	\draw [semithick,->] (15.6,6.7) to (16.2,6.7);
	\node at (16.5,6.75) {$x^{(\ell)}_{2K}$};
	
	\draw [fill] (2.7,8.25) circle [radius=0.025];
	\draw [fill] (2.7,8) circle [radius=0.025];
	\draw [fill] (2.7,7.75) circle [radius=0.025];
	
	\draw [fill] (7.05,7.8) circle [radius=0.025];
	\draw [fill] (7.05,7.55) circle [radius=0.025];
	\draw [fill] (7.05,7.3) circle [radius=0.025];
	
	\draw [fill] (8.2,7.5) circle [radius=0.025];
	\draw [fill] (8.2,7.25) circle [radius=0.025];
	\draw [fill] (8.2,7.0) circle [radius=0.025];

	\draw [fill] (11.5,8.1) circle [radius=0.025];
	\draw [fill] (11.5,7.85) circle [radius=0.025];
	\draw [fill] (11.5,7.6) circle [radius=0.025];

	\draw [fill] (12.7,7.65) circle [radius=0.025];
	\draw [fill] (12.7,7.4) circle [radius=0.025];
	\draw [fill] (12.7,7.15) circle [radius=0.025];

	\draw [fill] (13.7,7.65) circle [radius=0.025];
	\draw [fill] (13.7,7.4) circle [radius=0.025];
	\draw [fill] (13.7,7.15) circle [radius=0.025];

	\draw [fill] (16.4,8.1) circle [radius=0.025];
	\draw [fill] (16.4,7.85) circle [radius=0.025];
	\draw [fill] (16.4,7.6) circle [radius=0.025];
	\end{tikzpicture}
	\caption{Specific structure of layer $\ell$ of the proposed B-DetNet.}
	\label{fig_Bussgang_DetNet_layer_structure}
\end{figure*}
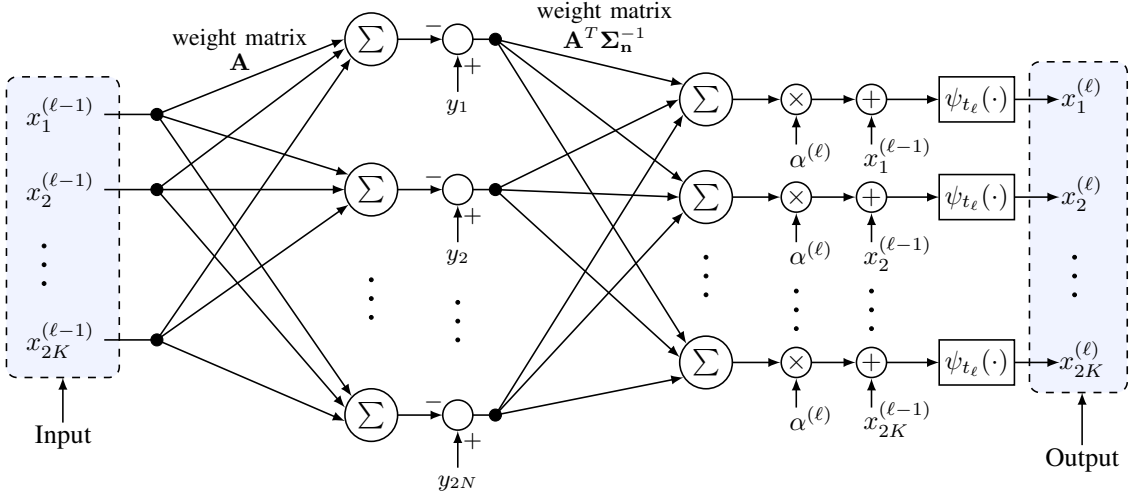

\subsection{Proposed B-DetNet}
Applying the Bussang decomposition to~\eqref{eq_quantized_real_received_signal}, we obtain
\begin{align}
    \mathbf{y} & = \mathbf{V}\mathbf{H}\mathbf{x} + \mathbf{V}\mathbf{z} + \mathbf{d}, \nonumber \\
    &=\mathbf{A}\mathbf{x} + \mathbf{n} \label{eq_linearized_data_system_model}
\end{align}
where 
\begin{align*}
    \mathbf{V} &= \frac{\Delta}{\sqrt{\pi}} \operatorname{diag}(\boldsymbol{\Sigma}_{\mathbf{r}})^{-\frac{1}{2}} \times \\ 
    & \qquad \sum_{i=1}^{2^b - 1}\exp \bigg \{-\Delta^2 (i - 2^{b-1})^2\operatorname{diag}(\boldsymbol{\Sigma}_{\mathbf{r}})^{-1} \bigg\}
\end{align*}
and $\boldsymbol{\Sigma}_{\mathbf{r}} = \frac{1}{2}\big(\mathbf{H}\mathbf{H}^T + N_0\mathbf{I}\big)$.

For the case of 1-bit ADCs, the covariance of $\mathbf{n}$ is given in a closed form as~\cite{mezghani2012capacity}
\begin{equation}\label{Sigma-n}
\begin{split}
\boldsymbol{\Sigma}_{\mathbf{n}} =& \frac{\Delta^2}{\pi}\Big[\operatorname{arcsin}\Big(\operatorname{diag}(\mathbf{\Sigma}_{\mathbf{r}})^{-\frac{1}{2}}\mathbf{\Sigma}_{\mathbf{r}}\operatorname{diag}(\mathbf{\Sigma}_{\mathbf{r}})^{-\frac{1}{2}}\Big)-\\
&\;\operatorname{diag}(\mathbf{\Sigma}_{\mathbf{r}})^{-\frac{1}{2}}\mathbf{\Sigma}_{\mathbf{r}}\operatorname{diag}(\mathbf{\Sigma}_{\mathbf{r}})^{-\frac{1}{2}}+ \frac{N_0}{2}\operatorname{diag}(\mathbf{\Sigma}_{\mathbf{r}})^{-1}\Big].
\end{split}
\end{equation}
For few-bit ADCs, the covariance of $\mathbf{n}$ can be approximated as $\boldsymbol{\Sigma}_{\mathbf{n}} \approx \frac{N_0}{2}\mathbf{V}\mathbf{V}^T + \eta_b\operatorname{diag}(\boldsymbol{\Sigma}_{\mathbf{r}})$. Here, the effective noise $\mathbf{n}$ is often modeled as Gaussian noise as $\mathcal{N}(\mathbf{0},\boldsymbol{\Sigma}_{\mathbf{n}})$. Based on this linearized model, different linear detectors for one-bit and few-bit ADCs such as BZF, BMMSE, and BWZF were introduced in~\cite{Lan2018Linearized,Ly2021Linear,Kolomvakis2020Quantized}.

Here, we propose a data detection network, namely B-DetNet, based on the linearized system model in~\eqref{eq_linearized_data_system_model}. Since the effective noise $\mathbf{n}$ is assumed to be Gaussian, the Bussgang-based maximum likelihood detection problem is given as
\begin{equation}
    \hat{\mathbf{x}}_{\mathtt{BML}} = \argmin_{\bar{\mathbf{x}}\in \bar{\mathcal{M}}^{K}}\; (\mathbf{y - Ax})^T\boldsymbol{\Sigma}^{-1}_{\mathbf{n}}(\mathbf{y-Ax}).
    \label{eq_BML_detction_problem}
\end{equation}
Let $P_{\mathrm{B}}(\mathbf{x})$ be the objective function of~\eqref{eq_BML_detction_problem}. Note that $P_{\mathrm{B}}(\mathbf{x})$ is a quadratic function of $\mathbf{x}$ and thus convex. However, the optimization problem is not convex due to the constraint on the discrete feasibility set $\bar{\mathcal{M}}^{K}$. An optimal solution to~\eqref{eq_BML_detction_problem} therefore requires an exhaustive search, which is very expensive for large scale systems. Instead, an iterative projected gradient descent method
\begin{equation}
    \mathbf{x}^{(\ell)} = \psi_{t_\ell}\left(\mathbf{x}^{(\ell-1)} - \alpha^{(\ell)} \nabla P_{\mathrm{B}}(\mathbf{x}^{(\ell-1)}) \right)
    \label{eq_Bussgang_projected_gradient_descent}
\end{equation}
can be applied to to search for its optimal solution. Herein, the gradient of $P_{\mathrm{B}}(\mathbf{x})$ evaluated at $\mathbf{x}^{(\ell-1)}$ is given by
\begin{equation}
    \nabla P_{\mathrm{B}}(\mathbf{x}^{(\ell-1)}) = -2\mathbf{A}^T\boldsymbol{\Sigma}^{-1}_{\mathbf{n}}\big(\mathbf{y} - \mathbf{A}\mathbf{x}^{(\ell-1)}\big)
    \label{eq_Bussgang_gradient}
\end{equation}
and $\psi_{t_\ell}(\cdot)$ characterized by a positive parameter $t_\ell$ is a non-linear projector to force the signal to the regime of constellation points. Based on the ReLU activation function, like $q(r)$ in~\eqref{eq_soft_quantizer_based_on_ReLU}, $\psi_{t_\ell}(\cdot)$ can be written as
\begin{align}
    \psi_{t_\ell}(x) & = -(2^{b'} - 1)\frac{\Delta'}{2} + \frac{\Delta'}{2t_\ell}  \sum_{i=-B'}^{B'}\big[f_{\mathrm{relu}}(r + i\Delta + t_\ell) -  \notag \\
    & \qquad f_{\mathrm{relu}}(r + i\Delta - t_\ell)\big]
    \label{eq_projector_based_on_ReLU}
\end{align}
where $B' = 2^{b'-1} - 1$ and $t_\ell$ is a positive number. For QPSK signalling, $\{b',\Delta'\} = \{1,\frac{2}{\sqrt{2}}\}$ and for $16$-QAM signalling, $\{b',\Delta'\} = \{2,\frac{2}{\sqrt{10}}\}$. Illustration for the effect of $t$ on $\psi_t(\cdot)$ is given in Fig.~\ref{fig_soft_projector}. It can also seen that smaller $t$ makes the projector sharper. Such a projection function was used in~\cite{Nhan2020Deep}, which studied deep learning-based detection for unquantized MIMO systems.

We propose B-DetNet by unfolding the projected gradient descent method in~\eqref{eq_Bussgang_projected_gradient_descent}. The overall structure of B-DetNet is illustrated in Fig.~\ref{fig_overall_network_structure}. There are $L$ layers where each layer takes an input vector of size $2K$ and generates an output vector of the same size. The specific layer structure of B-DetNet is given in Fig.~\ref{fig_Bussgang_DetNet_layer_structure} where $\mathbf{A}$ and $\mathbf{A}^T\boldsymbol{\Sigma}^{-1}_{\mathbf{n}}$ play the role of weight matrices. The received signal vector $\mathbf{y}$ can be seen as the bias vector. Hence, B-DetNet is highly adaptive to the channel. The only trainable parameters in a layer $\ell$ of B-DetNet are a step size $\alpha^{(\ell)}$ and a scaling parameter $t_{\ell}$ in the projector function $\psi_{t_\ell}(\cdot)$.

We note that similar structures for data detection in full-resolution systems have been developed in~\cite{Nhan2020Deep,Khani2020Adaptive}. However, the received signal in full-resolution systems is given as $\mathbf{y} = \mathbf{Hx+z}$, and therefore the gradient of interest is in the form of $-2\mathbf{H}^T(\mathbf{y-Hx})$. For low-resolution systems, we have a new effective channel $\mathbf{A}$ and a new noise covariance matrix $\boldsymbol{\Sigma}_{\mathbf{n}}$, resulting in a new form of gradient as in~\eqref{eq_Bussgang_gradient}.

\begin{figure*}[t!]
	\centering
	\begin{tikzpicture}
	\draw [fill=InOutColor, dashed, semithick, rounded corners] (-0.8,6.5) rectangle (0.6,10.5);
	\draw [semithick,->] (-0.05,5.9) to (-0.05,6.5);
	\node at (-0.05,5.7) {Input};
	
	\draw [fill=InOutColor, dashed, semithick, rounded corners] (15.8,6.3) rectangle (17.1,10.7);
	\draw [semithick,->] (16.5,5.6) to (16.5,6.3);
	\node at (16.5,5.4) {Output};
	
	\node at (2,11.2) {\small {weight matrix}};
	\node at (2,10.9) {\small {$\mathbf{H}$}};
	\node at (10.1,11.3) {\small {weight matrix}};
	\node at (10.1,11) {\small {$\mathbf{H}^T$}};
	
	\node [left] at (0.5,10) {$x^{(\ell-1)}_{1}$};
	\draw [semithick] (0.4,10) to (0.9,10);
	\draw [fill] (0.9,10) circle [radius=0.08];
	
	\draw [semithick, ->] (0.9,10) to (3.14,11);
	\draw [semithick, ->] (0.9,9) to (3.16,10.88);
	\draw [semithick, ->] (0.9,7) to (3.24,10.75);
	
	\draw [semithick] (3.5,11) circle [radius=0.35];
	\node at (3.5,11) {$\sum$};
	\draw [semithick] (3.85,11) to (4.7,11);
	\draw [semithick,<->] (4.99,11.6) to (4.7,11.6) to (4.7,10.4) to (4.99,10.4);
    \node at (4.3,11.3) {$u^{(\ell)}_{1}$};
	\draw [semithick] (5,11.3) rectangle (7.7,11.9);
	\node [right] at (5,11.6) {$\sigma\big(\beta(u^{(\ell)}_{1}-q^{\mathrm{up}}_1)\big)$};
	\draw [semithick] (5,10.1) rectangle (7.7,10.7);
	\node [right] at (4.96,10.4) {$\sigma\big(\beta(u^{(\ell)}_{1}-q^{\mathrm{low}}_1)\big)$};
	\draw [semithick,->] (7.7,11.6) to (8.2,11.6) to (8.2,11.21);
	\draw [semithick] (8.2,11) circle [radius=0.2];
	\draw [semithick,->] (7.7,10.4) to (8.2,10.4) to (8.2,10.79);
	\draw [semithick,->] (7.5,11) to (7.99,11);
	\node [left] at (7.5,11) {$1$};
	\node at (7.88,11.17) {{\small $+$}};
	\node at (8.4,11.35) {{\small $-$}};
	\node at (8.4,10.65) {{\small $-$}};
	\draw [semithick] (8.4,11) to (8.7,11);
	\draw [fill] (8.7,11) circle [radius=0.08];
	
	\draw [semithick,->] (8.7,11) to (11.2,10.4);
	\draw [semithick,->] (8.7,11) to (11.2,9.1);
	\draw [semithick,->] (8.7,11) to (11.24,6.95);
	
	\draw [semithick] (11.5,10.2) circle [radius=0.35];
	\node at (11.5,10.2) {$\sum$};
	\draw [semithick,->] (11.85,10.2) to (12.49,10.2);
	\draw [semithick] (12.7,10.2) circle [radius=0.2];
	\node at (12.7,10.2) {$\times$};
	\draw [semithick,->] (12.7,9.6) to (12.7,9.99);
	\node at (12.9,9.46) {$\alpha^{(\ell)}$};
	\draw [semithick,->] (12.9,10.2) to (13.49,10.2);
	\draw [semithick] (13.7,10.2) circle [radius=0.2];
	\node at (13.7,10.2) {$+$};
	\draw [semithick,->] (13.7,9.6) to (13.7,9.99);
	\node at (14.05,9.45) {$x^{(\ell-1)}_{1}$};
	\draw [semithick,->] (13.9,10.2) to (14.6,10.2);
	\draw [semithick] (14.6,9.9) rectangle (15.6,10.5);
	\node at (15.1,10.2) {$\psi_{t_\ell}(\cdot)$};
	\draw [semithick,->] (15.6,10.2) to (16.2,10.2);
	\node at (16.5,10.25) {$x^{(\ell)}_{1}$};
	
	\node [left] at (0.5,9) {$x^{(\ell-1)}_{2}$};
	\draw [semithick] (0.4,9) to (0.9,9);
	\draw [fill] (0.9,9) circle [radius=0.08];
	
	\draw [semithick, ->] (0.9,10) to (3.19,9.2);
	\draw [semithick, ->] (0.9,9) to (3.14,9);
	\draw [semithick, ->] (0.9,7) to (3.2,8.8);
	
	\draw [semithick] (3.5,9) circle [radius=0.35];
	\node at (3.5,9) {$\sum$};
	\draw [semithick] (3.85,9) to (4.7,9);
	\draw [semithick,<->] (4.99,9.6) to (4.7, 9.6) to (4.7,8.4) to (4.99,8.4);
    \node at (4.3,9.3) {$u^{(\ell)}_{2}$};
	\draw [semithick] (5,9.3) rectangle (7.7,9.9);
	\node [right] at (5,9.6) {$\sigma\big(\beta(u^{(\ell)}_{2}-q^{\mathrm{up}}_2)\big)$};
	\draw [semithick] (5,8.1) rectangle (7.7,8.7);
	\node [right] at (4.96,8.4) {$\sigma\big(\beta(u^{(\ell)}_{2}-q^{\mathrm{low}}_2)\big)$};
	%\draw [semithick] (7.7,9.6) to (8.2,9.6);
	\draw [semithick,->] (7.7,9.6) to (8.2,9.6) to (8.2,9.21);
	\draw [semithick] (8.2,9) circle [radius=0.2];
	\draw [semithick,->] (7.7,8.4) to (8.2,8.4) to (8.2,8.79);
	\draw [semithick,->] (7.5,9) to (7.99,9);
	\node [left] at (7.5,9) {$1$};
	\node at (7.88,9.17) {{\small $+$}};
	\node at (8.4,9.35) {{\small $-$}};
	\node at (8.4,8.65) {{\small $-$}};
	\draw [semithick] (8.4,9) to (8.7,9);
	\draw [fill] (8.7,9) circle [radius=0.08];
	
	\draw [semithick,->] (8.7,9) to (11.14,10.2);
	\draw [semithick,->] (8.7,9) to (11.15,8.9);
	\draw [semithick,->] (8.7,9) to (11.15,6.77);
	
	\draw [semithick] (11.5,8.9) circle [radius=0.35];
	\node at (11.5,8.9) {$\sum$};
	\draw [semithick,->] (11.85,8.9) to (12.49,8.9);
	\draw [semithick] (12.7,8.9) circle [radius=0.2];
	\node at (12.7,8.9) {$\times$};
	\draw [semithick,->] (12.7,8.3) to (12.7,8.69);
	\node at (12.9,8.16) {$\alpha^{(\ell)}$};
	\draw [semithick,->] (12.9,8.9) to (13.49,8.9);
	\draw [semithick] (13.7,8.9) circle [radius=0.2];
	\node at (13.7,8.9) {$+$};
	\draw [semithick,->] (13.7,8.3) to (13.7,8.69);
	\node at (14.05,8.15) {$x^{(\ell-1)}_{2}$};
	\draw [semithick,->] (13.9,8.9) to (14.6,8.9);
	\draw [semithick] (14.6,8.6) rectangle (15.6,9.2);
	\node at (15.1,8.9) {$\psi_{t_\ell}(\cdot)$};
	\draw [semithick,->] (15.6,8.9) to (16.2,8.9);
	\node at (16.5,8.95) {$x^{(\ell)}_{2}$};
	
	\node [left] at (0.5,7) {$x^{(\ell-1)}_{2K}$};
	\draw [semithick] (0.4,7) to (0.9,7);
	\draw [fill] (0.9,7) circle [radius=0.08];
	
	\draw [semithick, ->] (0.9,10) to (3.24,6.25);
	\draw [semithick, ->] (0.9,9) to (3.16,6.1);
	\draw [semithick, ->] (0.9,7) to (3.15,5.92);
	
	\draw [semithick] (3.5,6) circle [radius=0.35];
	\node at (3.5,6) {$\sum$};
	\draw [semithick] (3.85,6) to (4.7,6);
	\draw [semithick,<->] (4.99,6.6) to (4.7,6.6) to  (4.7,5.4) to (4.99,5.4);
    \node at (4.3,6.3) {$u^{(\ell)}_{2N}$};
	\draw [semithick] (5,6.3) rectangle (7.7,6.9);
	\node [right] at (4.96,6.6) {$\sigma\big(\beta(u^{(\ell)}_{2N}-q^{\mathrm{up}}_{2N})\big)$};
	\draw [semithick] (5,5.1) rectangle (7.7,5.7);
	\node [right] at (4.93,5.4) {$\sigma\big(\beta(u^{(\ell)}_{2N}-q^{\mathrm{low}}_{2N})\big)$};
	\draw [semithick,->] (7.7,6.6) to (8.2,6.6) to (8.2,6.21);
	\draw [semithick] (8.2,6) circle [radius=0.2];
	\draw [semithick,->] (7.7,5.4) to (8.2,5.4) to (8.2,5.79);
	\draw [semithick,->] (7.5,6) to (7.99,6);
	\node [left] at (7.5,6) {$1$};
	\node at (7.88,6.17) {{\small $+$}};
	\node at (8.4,6.35) {{\small $-$}};
	\node at (8.4,5.65) {{\small $-$}};
	\draw [semithick] (8.4,6) to (8.7,6);
	\draw [fill] (8.7,6) circle [radius=0.08];
	
	\draw [semithick,->] (8.7,6) to (11.2,10);
	\draw [semithick,->] (8.7,6) to (11.2,8.7);
	\draw [semithick,->] (8.7,6) to (11.2,6.5);
	
	\draw [semithick] (11.5,6.7) circle [radius=0.35];
	\node at (11.5,6.7) {$\sum$};
	\draw [semithick,->] (11.85,6.7) to (12.49,6.7);
	\draw [semithick] (12.7,6.7) circle [radius=0.2];
	\node at (12.7,6.7) {$\times$};
	\draw [semithick,->] (12.7,6.1) to (12.7,6.49);
	\node at (12.9,5.96) {$\alpha^{(\ell)}$};
	\draw [semithick,->] (12.9,6.7) to (13.49,6.7);
	\draw [semithick] (13.7,6.7) circle [radius=0.2];
	\node at (13.7,6.7) {$+$};
	\draw [semithick,->] (13.7,6.1) to (13.7,6.49);
	\node at (14.05,5.95) {$x^{(\ell-1)}_{2K}$};
	\draw [semithick,->] (13.9,6.7) to (14.6,6.7);
	\draw [semithick] (14.6,6.4) rectangle (15.6,7.0);
	\node at (15.1,6.7) {$\psi_{t_\ell}(\cdot)$};
	\draw [semithick,->] (15.6,6.7) to (16.2,6.7);
	\node at (16.5,6.75) {$x^{(\ell)}_{2K}$};
	
	%\draw [semithick,->] (13.9,6.7) to (15,6.7);
	%\node at (15.35,6.75) {$x^{(\ell)}_{2K}$};
	
	\draw [fill] (-0.4,8.25) circle [radius=0.025];
	\draw [fill] (-0.4,8) circle [radius=0.025];
	\draw [fill] (-0.4,7.75) circle [radius=0.025];
	
	\draw [fill] (3.6,8) circle [radius=0.025];
	\draw [fill] (3.6,7.75) circle [radius=0.025];
	\draw [fill] (3.6,7.5) circle [radius=0.025];
	
	\draw [fill] (6.3,7.75) circle [radius=0.025];
	\draw [fill] (6.3,7.5) circle [radius=0.025];
	\draw [fill] (6.3,7.25) circle [radius=0.025];

	\draw [fill] (11.5,8.1) circle [radius=0.025];
	\draw [fill] (11.5,7.85) circle [radius=0.025];
	\draw [fill] (11.5,7.6) circle [radius=0.025];

	\draw [fill] (12.7,7.65) circle [radius=0.025];
	\draw [fill] (12.7,7.4) circle [radius=0.025];
	\draw [fill] (12.7,7.15) circle [radius=0.025];

	\draw [fill] (13.7,7.65) circle [radius=0.025];
	\draw [fill] (13.7,7.4) circle [radius=0.025];
	\draw [fill] (13.7,7.15) circle [radius=0.025];

	\draw [fill] (16.4,8.1) circle [radius=0.025];
	\draw [fill] (16.4,7.85) circle [radius=0.025];
	\draw [fill] (16.4,7.6) circle [radius=0.025];
	\end{tikzpicture}
	\caption{Specific structure of layer $\ell$ of FBM-DetNet. The weight matrices and the bias vectors are defined by the channel and the received signal, respectively.}
	\label{fig_FBM_DetNet_layer_structure}
\end{figure*}
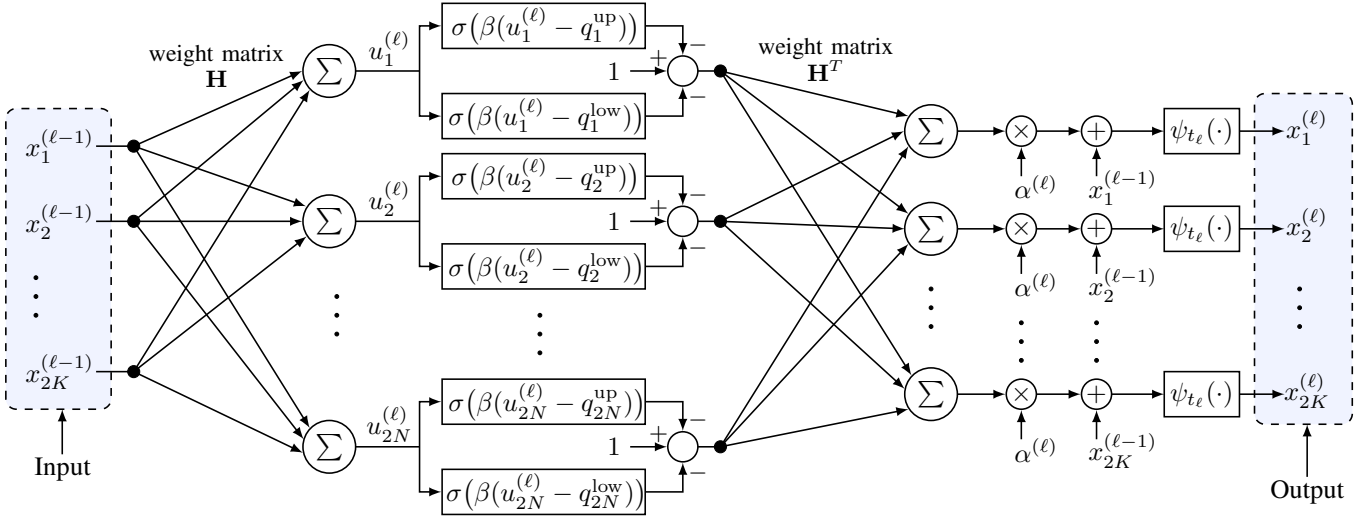
\subsection{Proposed FBM-DetNet}
\subsubsection{Maximum-likelihood data detection problem}
Let $s^\mathrm{up}_i=\sqrt{2\rho}(q^{\mathrm{up}}_i-\mathbf{h}_i^T\mathbf{x})$ and $s^\mathrm{low}_i=\sqrt{2\rho}(q^{\mathrm{low}}_i-\mathbf{h}_i^T\mathbf{x})$, where
\begin{align*}
    q^{\mathrm{up}}_i &= 
    \begin{cases}
    y_i+\frac{\Delta}{2} & \text{if}\; y_i<\tau_{2^b-1}\\
    \infty & \text{otherwise},
    \end{cases}\\
    q^{\mathrm{low}}_i &= 
    \begin{cases}
    y_i-\frac{\Delta}{2} & \text{if}\; y_i>\tau_1\\
    -\infty & \text{otherwise}.
    \end{cases}
\end{align*}
Hence, $q^{\mathrm{up}}_i$ and $q^{\mathrm{low}}_i$ are the upper and lower quantization thresholds of the bin to which $y_i$ belongs. The ML detection problem based on the log-likelihood function for the model in~\eqref{eq_quantized_real_received_signal} is defined as follows~\cite{Mezghani2008Maximum}:
\begin{eqnarray}
\hat{\mathbf{x}}_{\mathtt{ML}}
&=&\arg\max_{\bar{\mathbf{x}}\in \bar{\mathcal{M}}^{K}}\; \sum_{i=1}^{2N}\log\left[\Phi \left(s^\mathrm{up}_i\right)-\Phi \left(s^\mathrm{low}_i\right)\right]. 
\label{eq_conventional_logML_detection}
\end{eqnarray}
Let $\mathcal{P}(\mathbf{x})$ denote the objective function of~\eqref{eq_conventional_logML_detection}, which is a concave function of $\mathbf{x}$. However, the optimization problem~\eqref{eq_conventional_logML_detection} is not convex since the feasible set is a discrete set. Therefore, an optimal solution for ML detection in~\eqref{eq_conventional_logML_detection} also requires an exhaustive search over $\bar{\mathcal{M}}^{K}$, which is probihitively complex for large-scale systems. One can relax the constraint on the feasible set from $\bar{\mathbf{x}}\in \bar{\mathcal{M}}^{K}$ to $\bar{\mathbf{x}}\in \mathbb{C}^{K}$ in order to obtain a convex optimization problem and thus an iterative gradient descent method can be used. Unfortunately, such a method also suffers from the vanishing gradient issue as presented in the channel estimation problem. In addition, there is no closed-form expression for $\Phi(\cdot)$, which complicates the evaluation in \eqref{eq_conventional_logML_detection}. Thus, we also exploit the approximation in~\eqref{eq_approximate_Phi_as_Sigma} to obtain an approximate version of the function $\mathcal{P}(\mathbf{x})$ as follows:
\begin{align}
    \mathcal{P}(\mathbf{x}) & \approx \tilde{\mathcal{P}}(\mathbf{x}) = \sum_{i=1}^{2N}\log\left[\frac{1}{1+e^{-cs^\mathrm{up}_i}}-\frac{1}{1+e^{-cs^\mathrm{low}_i}}\right]\\
    &= \sum_{i=1}^{2N}\Big[\log\Big(e^{-cs^\mathrm{low}_i}-e^{-cs^\mathrm{up}_i}\Big) - \notag\\ & \qquad \quad \;\; \log\left(1+e^{-cs^\mathrm{up}_i}\right)- \log\left(1+e^{-cs^\mathrm{low}_i}\right)\Big].
\end{align}

The reformulated ML detection problem is thus
\begin{equation}
\hat{\mathbf{x}}_{\mathtt{ML}} = \arg\max_{\bar{\mathbf{x}}\in \bar{\mathcal{M}}^{K}}\; \tilde{\mathcal{P}}(\mathbf{x}).
\label{eq_reformulated_logML_detection}
\end{equation}
%Note that an optimal solution to problem \eqref{eq_reformulated_logML_detection} still requires an exhaustive search over $\bar{\mathcal{M}}^{K}$. Thus, we relax the constraint $\bar{\mathbf{x}}\in \bar{\mathcal{M}}^{K}$ in~\eqref{eq_reformulated_logML_detection} to $\bar{\mathbf{x}}\in \mathbb{C}^{K}$ and solve the following optimization problem:
%\begin{equation}
%\begin{aligned}
%& \underset{\bar{\mathbf{x}}\in \mathbb{C}^{K}}{\operatorname{maximize}}
%& & \tilde{\mathcal{P}}(\mathbf{x}).
%\end{aligned}
%\label{eq_relaxed_logML_detection}
%\end{equation}
The gradient of $\tilde{\mathcal{P}}(\mathbf{x})$ is 
\begin{align}
    \nabla\tilde{\mathcal{P}}(\mathbf{x})&=\sum_{i=1}^{2N}c\sqrt{2\rho}\,\mathbf{h}_i\left(1-\frac{1}{1+e^{cs^\mathrm{up}_i}}-\frac{1}{1+e^{cs^\mathrm{low}_i}}\right)\\
    &= c\sqrt{2\rho}\,\mathbf{H}^T\Big[\mathbf{1}-\sigma\left(c\sqrt{2\rho}\left(\mathbf{H}\mathbf{x} - \mathbf{q}^{\mathrm{up}}\right)\right)-\notag\\
    &\qquad\qquad\qquad\qquad\sigma\left(c\sqrt{2\rho}\left(\mathbf{H}\mathbf{x}-\mathbf{q}^{\mathrm{low}}\right)\right)\Big]
\end{align}
where $\mathbf{q}^{\mathrm{up}} = [q_1^{\mathrm{up}},\ldots,q_{2N}^{\mathrm{up}}]^T$ and $\mathbf{q}^{\mathrm{low}} = [q_1^{\mathrm{low}},\ldots,q_{2N}^{\mathrm{low}}]^T$. Thus, an iterative projected gradient decent method for solving~\eqref{eq_reformulated_logML_detection} can be written as
\begin{equation}
    \mathbf{x}^{(\ell)} = \psi_{t_\ell}\left(\mathbf{x}^{(\ell-1)} + \alpha^{(\ell)} \nabla\tilde{\mathcal{P}}(\mathbf{x}^{(\ell-1)})\right)
    \label{eq_iterative_method}
\end{equation}
where $\ell$ is the iteration index, $\alpha^{(\ell)}$ is a step size, and $\psi_{t_\ell}(\cdot)$ is also a projector as defined in~\eqref{eq_projector_based_on_ReLU}.

\subsubsection{Network structure of the proposed FBM-DetNet}
In order to optimize the step sizes $\{\alpha^{(\ell)}\}$ and scaling parameters $\{t_\ell\}$ of the projection function, we also use the deep unfolding technique \cite{Hershey-Unfolding-2014} to unfold each iteration in~\eqref{eq_iterative_method} as a layer of a DNN. The overall structure of the proposed DNN-based data detector is also illustrated in Fig.~\ref{fig_overall_network_structure}. The overall structure of FBM-DetNet is similar to that of B-DetNet as each layer of both the networks takes a vector of $2K$ elements as the input and generates an output vector of the same size. 

The specific structure for each layer $\ell$ of the proposed FBM-DetNet is illustrated in Fig.~\ref{fig_FBM_DetNet_layer_structure}. Each layer of FBM-DetNet has two weight matrices $\mathbf{H}$ and $\mathbf{H}^T$, and two bias vectors $\mathbf{q}^{\mathrm{up}}$ and $\mathbf{q}^{\mathrm{low}}$. These weight matrices and bias vectors are defined by the channel and the received signal, respectively. The activation function is the Sigmoid function $\sigma(\cdot)$ due to the use of the approximation in~\eqref{eq_approximate_Phi_as_Sigma}. Since $\mathbf{H} \in \mathbb{R}^{2N\times2K}$, the learning process for each layer of the proposed FBM-DetNet can be interpreted as first up-converting the signal $\mathbf{x}^{(\ell-1)}$ from dimension $2K$ to dimension $2N$ using the weight matrix $\mathbf{H}$, then applying nonlinear activation functions $\sigma(\cdot)$ before down-converting the signal back to dimension $2K$ using the weight matrix $\mathbf{H}^T$. Finally, the function $\psi_{t_\ell}(\cdot)$ is implemented to project $\mathbf{x}^{(\ell-1)}$ into the discrete set $\bar{\mc{M}}^K$.

It is observed that the layer structure of FBM-DetNet is similar to that of FBM-CENet in Fig.~\ref{fig_proposed_chanest_layer_structure}. However, while the weight matrices of FBM-CENet are defined by the pilot matrix $\mathbf{P}$ which is trainable, the weight matrices of FBM-DetNet are defined by the channel matrix $\mathbf{H}$ and thus not trainable. In other words, FBM-DetNet is highly adaptive to the channel. The trainable parameters of FBM-DetNet are the step sizes $\{\alpha^{(\ell)}\}$, scaling parameters $\{t_{\ell}\}$ for the projector, and a scaling parameter $\beta$ for the Sigmoid function. Note that the coefficient $c\sqrt{2\rho}$ is also omitted in FBM-DetNet for the same reason as in FBM-CENet. %It should also be noted that the layer structures of FBM-DetNet in Fig.~\ref{fig_FBM_DetNet_layer_structure} and B-DetNet in Fig.~\ref{fig_Bussgang_DetNet_layer_structure} are fully connected whereas the connections in each layer of FBM-CENet in Fig.~\ref{fig_proposed_chanest_layer_structure} are sparse since the pilot matrix $\bar{\mathbf{P}} = \bar{\mathbf{X}}_\mathrm{t}^T\otimes\mathbf{I}_N$ is a block-diagonal matrix.

\subsection{Training strategy}
A training sample for the two proposed data detection networks, B-DetNet and FBM-DetNet, can be obtained by randomly generating a channel matrix $\mathbf{H}$, a transmitted signal $\mathbf{x}$, and a noise vector $\mathbf{z}$. The cost function to be minimized is $\|\mathbf{x}^{(L)}-\mathbf{x}\|^2$, where $\mathbf{x}$ is the target signal, i.e., the transmitted signal. For training the proposed data detection networks, we do not need to use the soft quantization model because the trainable parameters do not appear in the received signals $\mathbf{y}$ or $\mathbf{q}^{\mathrm{up}}$ and $\mathbf{q}^{\mathrm{low}}$. These received signals are defined given a training sample $\{\mathbf{H}, \mathbf{x}, \mathbf{z}\}$, and therefore the hard quantizer can be used.

%\section{Pilot Signal Design}

%\section{Constellation Design}
%Consider the mapping from bits to symbols. Set individual encoder for each user, where the mapping is from $m$ bits to $2$-dimensional signal. At the receiver, we use the demapping from each estimated symbol back to the $m$ bits. A sigmoid function is applied at each output. This allows the bit-wise mapping and demapping. It is possible that at high modulation scheme (16-QAM), the mapping may come out better than the Gray mapping with square QAM.

\begin{figure}[t!]
    \centering
    \includegraphics[width=0.5\textwidth]{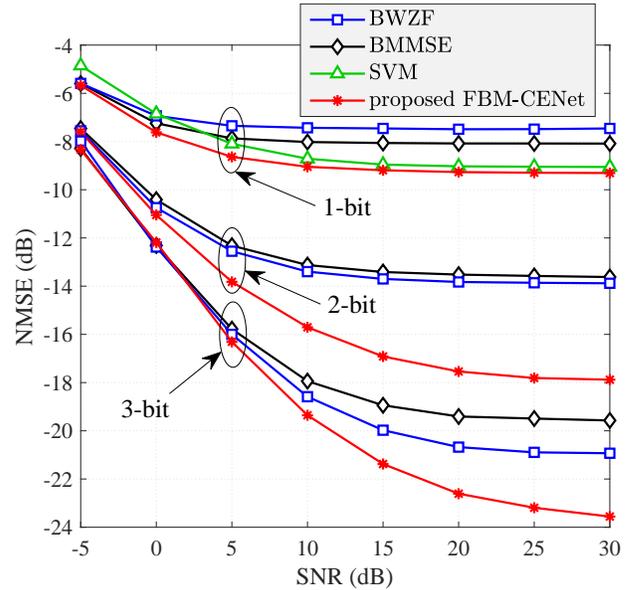}
    \caption{Channel estimation performance comparison for a given pilot matrix with $K = 4$ and $L = 8$.}
    \label{fig_NMSE_known_pilot}
\end{figure}
\section{Numerical Results}
\label{sec_numerical_results}
This section presents numerical results to show the superiority of the proposed channel estimation and data detection networks. The channel elements are assumed to be i.i.d. and each channel element is generated from the normal distribution $\mathcal{CN}(0, 1)$. 

For training the networks, we use TensorFlow~\cite{tensorflow} and the Adam optimizer~\cite{kingma2014adam} with a learning rate starts at $0.002$ and decays at a rate of $0.97$ after every $100$ training epochs. The size of each training batch is set to $1000$. The input of the first layer is set to a zero vector. In case the pilot matrix $\mathbf{P}$ is trainable, we use the soft quantization model in~\eqref{eq_soft_qup_based_on_relu} and~\eqref{eq_soft_qlow_based_on_relu} for the training phase and set $c_1 = 0.01$ and $c_2 = 1000$. For the channel estimation phase, we set the training length to be five times the number of users, i.e., $T_\mathrm{t} = 5K$.

Fig.~\ref{fig_NMSE_known_pilot} presents a performance comparison of different channel estimation methods for a given pilot matrix in terms of NMSE, defined here as $\mathrm{NMSE} = \mathbb{E}[\|\hat{\mathbf{H}} - \bar{\mathbf{H}}\|_{\mathrm{F}}^2]/(KN)$, where $\hat{\mathbf{H}}$ is a estimate of the channel $\bar{\mathbf{H}}$. The given pilot matrix contains $K$ columns of a discrete Fourier transform (DFT) matrix where the $k^{\text{th}}$ row of the pilot matrix $\bar{\mathbf{X}}_\mathrm{t}$ is the $(k+1)^\text{th}$ column of the DFT matrix of size $T_\mathrm{t}\times T_\mathrm{t}$. In case of one-bit ADCs, it is observed that the proposed FBM-CENet slightly outperforms the SVM-based method in~\cite{Ly2021SVM} at medium-to-high SNRs. However, at low SNRs, the performance gap between the proposed FBM-CENet and the SVM method is larger since the SVM method does not perform well at low SNRs. For few-bit ADCs, it is clear to see that the proposed FBM-CENet significantly outperforms other existing channel estimation methods. Note that the SVM-based method in~\cite{Ly2021SVM} was specifically developed for one-bit ADCs. Therefore, we do not have results of the SVM-based method for few-bit ADCs. Note that the BWZF method does not perform well in case of one-bit ADCs because the BWZF method exploits the fact that the variance of the received signals at different quantization bins are different and sets the signals with lower variance to have higher weight. However, in case of one-bit ADCs, there is only one bin in each quantization side (positive or negative side). Therefore, there is no weight effects for one-bit ADCs. On the other hand, more quantization bits result in more quantization bins and thus different weights come into play. In other words, BWZF performs better with few-bit quantization.

\begin{figure}[t!]
    \centering
    \includegraphics[width=0.5\textwidth]{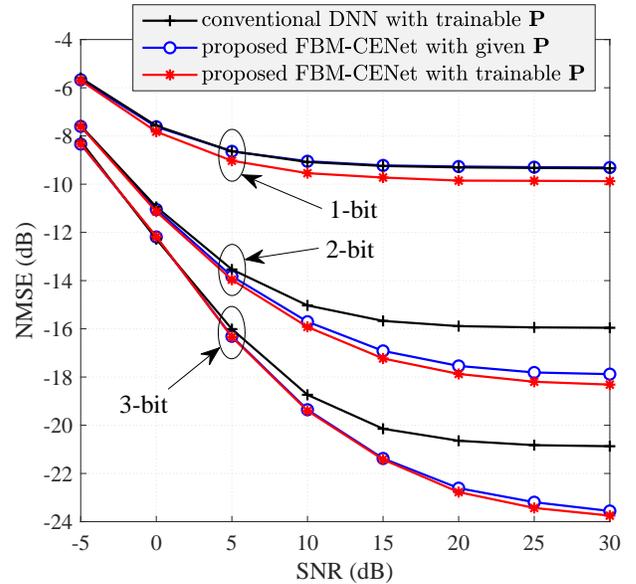}
    \caption{Channel estimation performance comparison with trainable pilot matrix, $K = 4$ and $L = 8$.}
    \label{fig_NMSE_trainable_pilot}
\end{figure}

In Fig.~\ref{fig_NMSE_trainable_pilot}, we consider the case where the pilot matrix is trained concurrently with the channel estimator. The proposed FBM-CENet is compared with an existing conventional DNN-based method in~\cite{DuyNguyen2020Neural} which also jointly optimizes the pilot matrix and the channel estimator like the proposed FBM-CENet. It can be seen the proposed FBM-CENet significantly outperforms the channel estimator in~\cite{DuyNguyen2020Neural}. The reason is that the estimation network in~\cite{DuyNguyen2020Neural} uses the data-driven conventional DNN structure as illustrated in Fig.~\ref{fig_conventional_DNN_layer_structure}. On the other hand, the structure of the proposed FBM-CENet takes advantages of the domain knowledge in the ML estimation framework. In Fig.~\ref{fig_NMSE_trainable_pilot}, we also include the channel estimation performance of FBM-CENet for a given pilot matrix in order to show that jointly optimizing the pilot matrix and the estimator can improve the estimation accuracy.

\begin{figure*}[t!]
	\centering
	\begin{subfigure}[t]{0.32\textwidth}
		\centering
		\includegraphics[width=\linewidth]{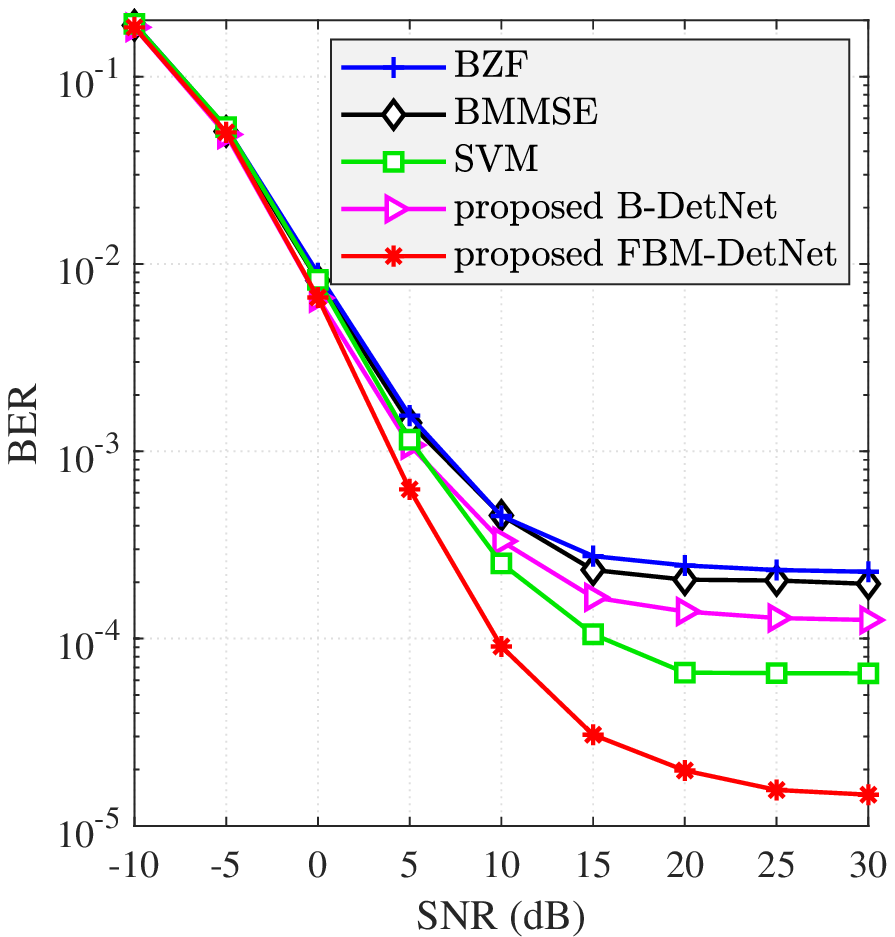}
		\caption{$b=1$ bit, $K=4$, and $L = 8$.}
		\label{fig_BER_QPSK_1bit_4K_32N}
	\end{subfigure}~
	\begin{subfigure}[t]{0.32\textwidth}
		\centering
		\includegraphics[width=\linewidth]{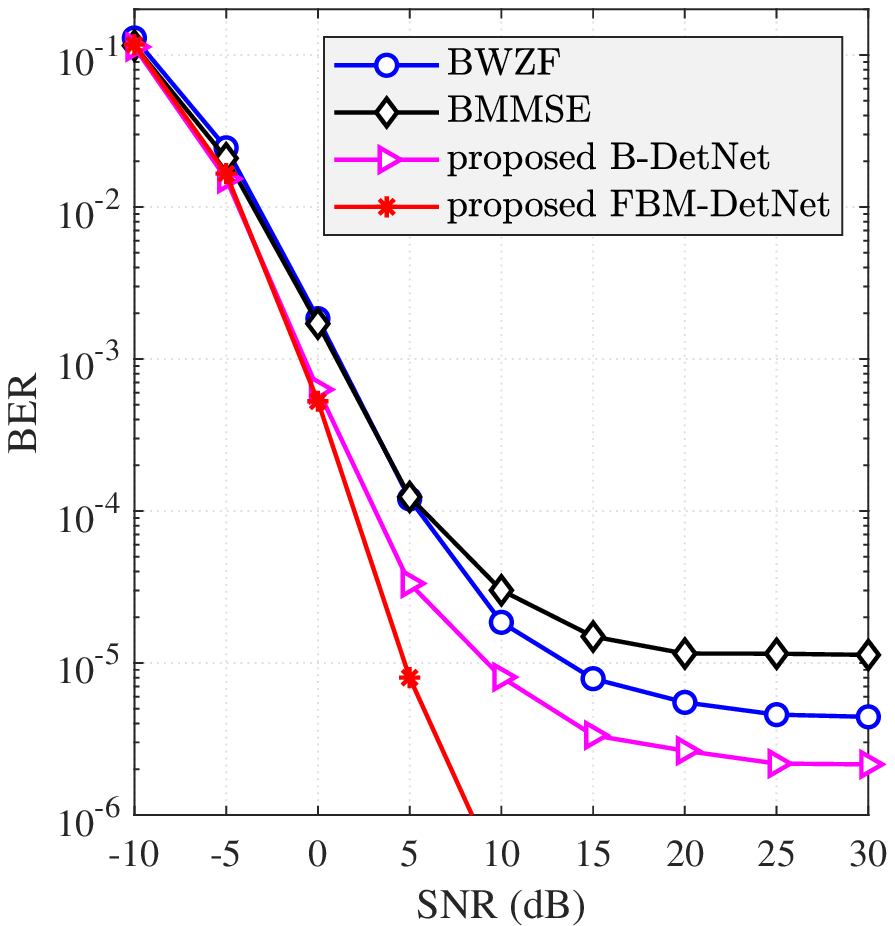}
		\caption{$b=2$ bit, $K=8$, and $L = 16$.}
		\label{fig_BER_QPSK_2bit_8K_32N}
	\end{subfigure}~
	\begin{subfigure}[t]{0.32\textwidth}
		\centering
		\includegraphics[width=\linewidth]{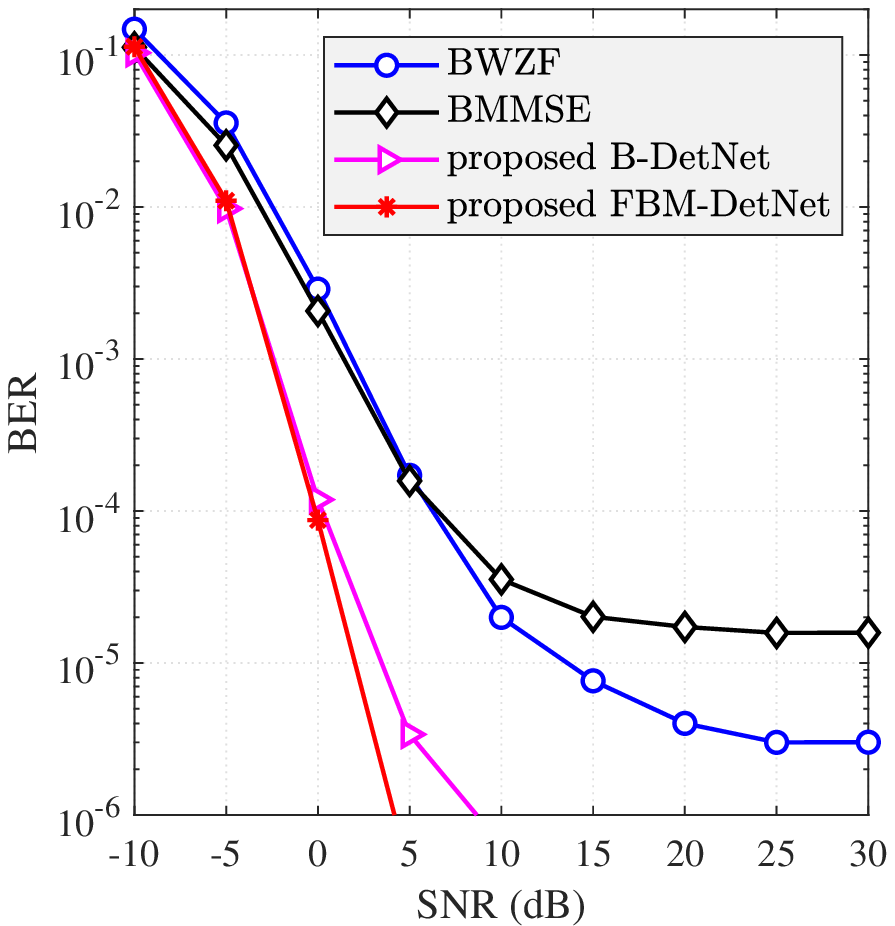}
		\caption{$b=3$ bit, $K=16$, and $L = 24$.}
		\label{fig_BER_QPSK_3bit_16K_32N}
	\end{subfigure}
	\caption{Performance comparison for data detection methods with QPSK signalling and $N = 32$.}
	\label{fig_BER_QPSK_32N}
\end{figure*}

\begin{figure*}[t!]
	\centering
	\begin{subfigure}[t]{0.32\textwidth}
		\centering
		\includegraphics[width=\linewidth]{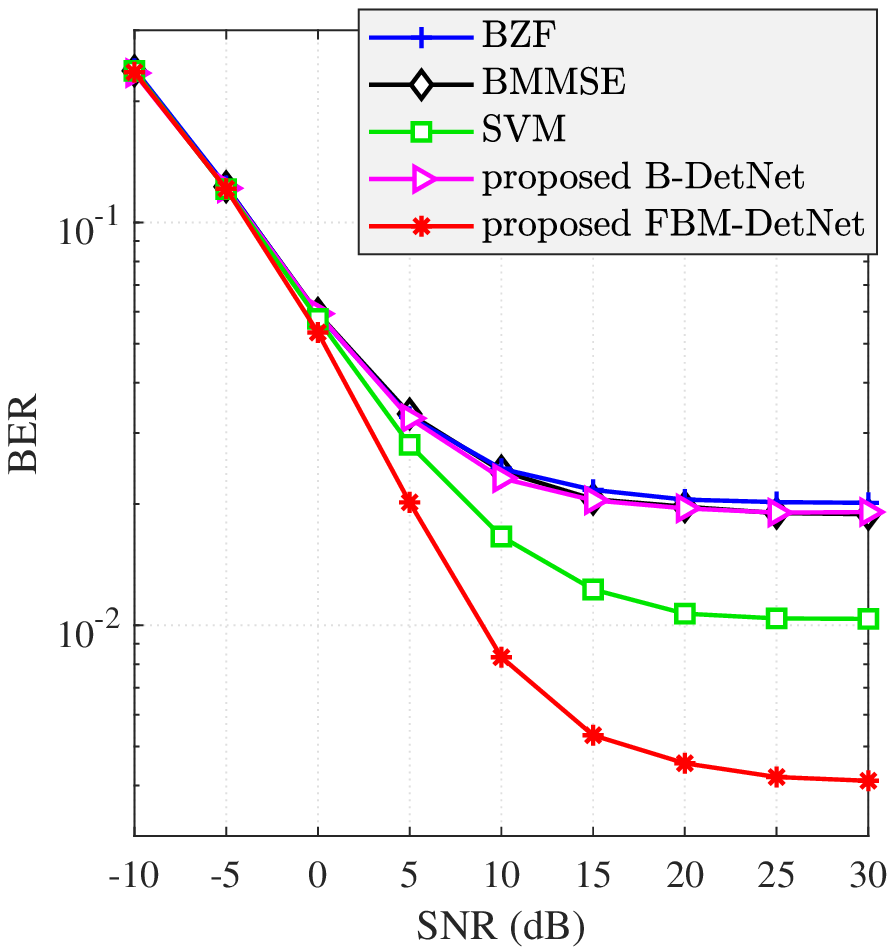}
		\caption{$b=1$ bit, $K=4$, and $L = 8$.}
		\label{fig_BER_16QAM_1bit_4K_64N}
	\end{subfigure}~
	\begin{subfigure}[t]{0.32\textwidth}
		\centering
		\includegraphics[width=\linewidth]{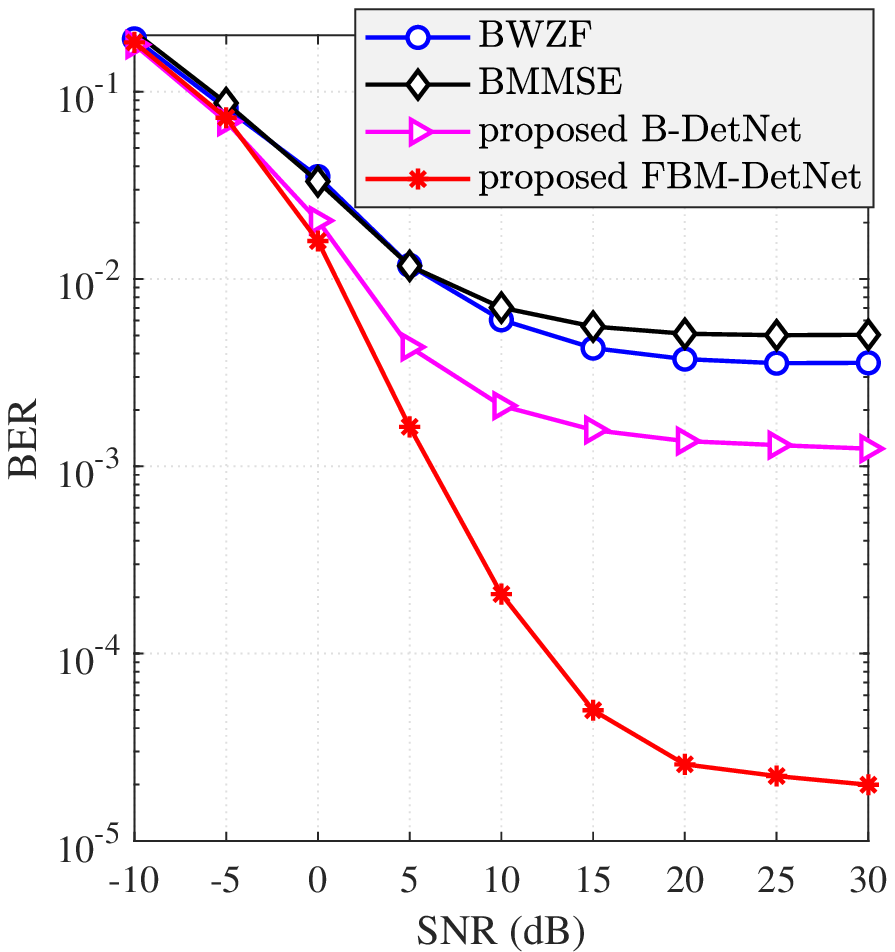}
		\caption{$b=2$ bit, $K=8$, and $L = 16$.}
		\label{fig_BER_16QAM_2bit_8K_64N}
	\end{subfigure}~
	\begin{subfigure}[t]{0.32\textwidth}
		\centering
		\includegraphics[width=\linewidth]{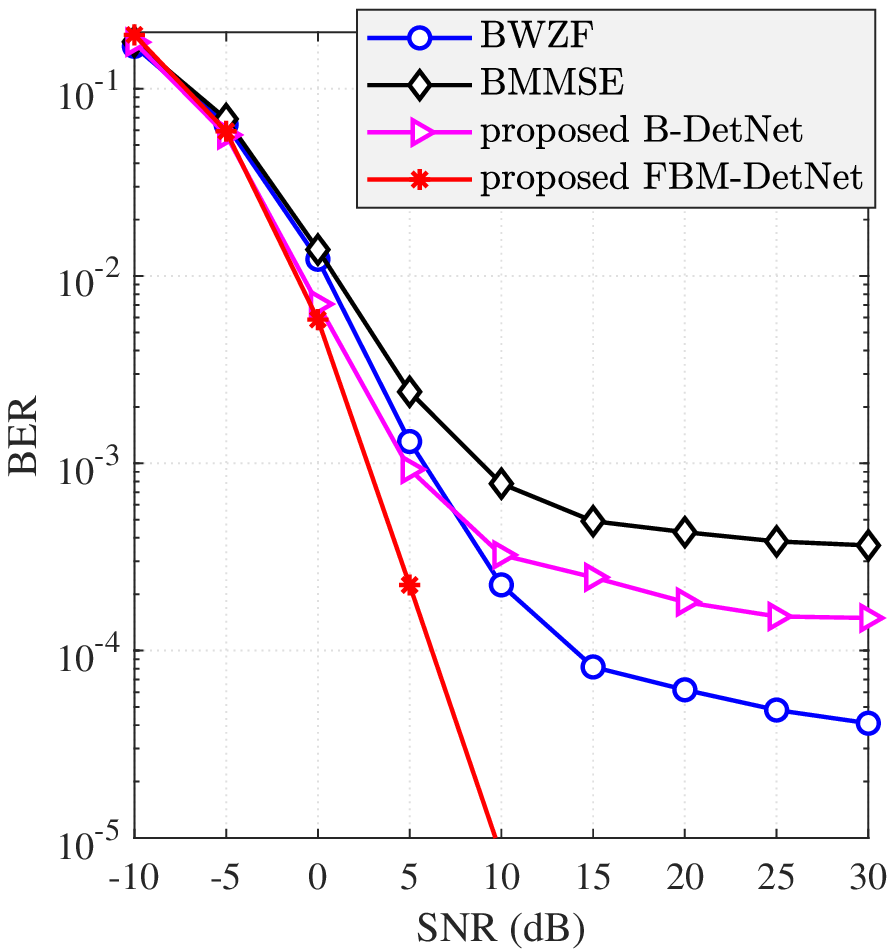}
		\caption{$b=3$ bit, $K=16$, and $L = 24$.}
		\label{fig_BER_16QAM_3bit_16K_64N}
	\end{subfigure}
	\caption{Performance comparison for data detection methods with $16$QAM signalling and $N = 64$.}
	\label{fig_BER_16QAM_64N}
\end{figure*}

Performance comparison for data detection is given in Fig.~\ref{fig_BER_QPSK_32N} and Fig.~\ref{fig_BER_16QAM_64N} for QPSK signalling and $16$QAM signalling, respectively. In these figures, we use the estimated CSI obtained by the proposed FBM-CENet with trainable pilot matrix. It can be easily seen that the proposed FBM-DetNet significantly outperforms other data detection methods. We note that B-DetNet performs worse than FBM-DetNet because FBM-DetNet is developed based on the original quantized system model whereas B-DetNet relies on a linearized system model in~\eqref{eq_linearized_data_system_model} whose the effective noise $\mathbf{n}$ is assumed to be Gaussian for simplicity but in fact $\mathbf{n}$ is not Gaussian. Furthermore, the distortion covariance matrix $\boldsymbol{\Sigma}_{\mathbf{n}}$ for the case of few-bit ADCs is approximate since a closed-form expression of $\boldsymbol{\Sigma}_{\mathbf{n}}$ is intractable.

\section{Conclusion}
\label{sec_conclusion}
In this paper, we have developed a channel estimation network (FBM-CENet) and two data detection networks (B-DetNet and FBM-DetNet) for massive MIMO systems with low-resolution ADCs. The proposed networks are model-driven and have special structures that can take advantages of domain-knowledge to efficiently address the severe non-linearity caused by the low-resolution ADCs. An interesting feature of the proposed FBM-CENet is that the pilot matrix directly plays the role of the weight matrices in the network structure. Such a feature makes it possible to jointly optimize the estimation network and the pilot signal by simply treating the pilot matrix as trainable parameters. The proposed detection networks are highly adaptive to the channel and easy to train since they have a small number of trainable parameters in the network structures. Simulation results have shown that the proposed networks significantly outperform existing methods.

\ifCLASSOPTIONcaptionsoff
  \newpage
\fi

\bibliographystyle{IEEEtran}
\bibliography{main.bib}

\begin{thebibliography}{10}
\providecommand{\url}[1]{#1}
\csname url@samestyle\endcsname
\providecommand{\newblock}{\relax}
\providecommand{\bibinfo}[2]{#2}
\providecommand{\BIBentrySTDinterwordspacing}{\spaceskip=0pt\relax}
\providecommand{\BIBentryALTinterwordstretchfactor}{4}
\providecommand{\BIBentryALTinterwordspacing}{\spaceskip=\fontdimen2\font plus
\BIBentryALTinterwordstretchfactor\fontdimen3\font minus
  \fontdimen4\font\relax}
\providecommand{\BIBforeignlanguage}[2]{{%
\expandafter\ifx\csname l@#1\endcsname\relax
\typeout{** WARNING: IEEEtran.bst: No hyphenation pattern has been}%
\typeout{** loaded for the language `#1'. Using the pattern for}%
\typeout{** the default language instead.}%
\else
\language=\csname l@#1\endcsname
\fi
#2}}
\providecommand{\BIBdecl}{\relax}
\BIBdecl

\bibitem{walden1999analog}
R.~H. Walden, ``{A}nalog-to-digital converter survey and analysis,'' \emph{IEEE
  J. Select. Areas in Commun.}, vol.~17, no.~4, pp. 539--550, Apr. 1999.

\bibitem{choi2016near}
J.~Choi, J.~Mo, and R.~W. Heath, ``{N}ear maximum-likelihood detector and
  channel estimator for uplink multiuser massive {MIMO} systems with one-bit
  {ADC}s,'' \emph{IEEE Trans. Commun.}, vol.~64, no.~5, pp. 2005--2018, May
  2016.

\bibitem{li2017channel}
Y.~{Li}, C.~{Tao}, G.~{Seco-Granados}, A.~{Mezghani}, A.~L. {Swindlehurst}, and
  L.~{Liu}, ``Channel estimation and performance analysis of one-bit massive
  {MIMO} systems,'' \emph{IEEE Trans. Signal Process.}, vol.~65, no.~15, pp.
  4075--4089, Aug. 2017.

\bibitem{Shilpa2019Massive}
S.~{Rao}, A.~L. {Swindlehurst}, and H.~{Pirzadeh}, ``Massive {MIMO} channel
  estimation with 1-bit spatial sigma-delta {ADC}s,'' in \emph{Proc. IEEE Int.
  Conf. on Acoustics, Speech and Signal Process.}, Brighton, United Kingdom,
  May 2019, pp. 4484--4488.

\bibitem{Zhichao2019Oversampling}
Z.~{Shao}, L.~T.~N. {Landau}, and R.~C. d.~{Lamare}, ``Oversampling based
  channel estimation for 1-bit large-scale multiple-antenna systems,'' in
  \emph{Proc. Int. ITG Workshop on Smart Antennas}, Vienna, Austria, April
  2019.

\bibitem{Zhichao2019Channel}
Z.~{Shao}, L.~T.~N. {Landau}, and R.~C. {de Lamare}, ``Channel estimation using
  1-bit quantization and oversampling for large-scale multiple-antenna
  systems,'' in \emph{Proc. IEEE Int. Conf. on Acoustics, Speech and Signal
  Process.}, Brighton, United Kingdom, May 2019, pp. 4669--4673.

\bibitem{Liu2020Angular}
F.~{Liu}, H.~{Zhu}, C.~{Li}, J.~{Li}, P.~{Wang}, and P.~{Orlik},
  ``Angular-{D}omain channel estimation for one-bit massive {MIMO} systems:
  {P}erformance bounds and algorithms,'' \emph{IEEE Trans. Veh. Technol.},
  vol.~69, no.~3, pp. 2928--2942, Mar. 2020.

\bibitem{Kim2018Dominant}
I.~{Kim}, N.~{Lee}, and J.~{Choi}, ``Dominant channel estimation via {MIPS} for
  large-scale antenna systems with one-bit {ADC}s,'' in \emph{Proc. IEEE Global
  Commun. Conf.}, Abu Dhabi, United Arab Emirates, Dec. 2018.

\bibitem{Kim2018Channel}
H.~{Kim} and J.~{Choi}, ``Channel {A}o{A} estimation for massive {MIMO} systems
  using one-bit {ADC}s,'' \emph{Journal of Communications and Networks},
  vol.~20, no.~4, pp. 374--382, Aug. 2018.

\bibitem{kim2019channel2}
H.~Kim and J.~Choi, ``Channel estimation for spatially/temporally correlated
  massive {MIMO} systems with one-bit {ADC}s,'' \emph{EURASIP J. Wireless
  Commun. and Networking}, vol. 2019, no.~1, p. 267, 2019.

\bibitem{Srinivas2019Itervative}
B.~{Srinivas}, K.~{Mawatwal}, D.~{Sen}, and S.~{Chakrabarti}, ``An iterative
  semi-blind channel estimation scheme and uplink spectral efficiency of pilot
  contaminated one-bit massive {MIMO} systems,'' \emph{IEEE Tran. Veh.
  Technol.}, vol.~68, no.~8, pp. 7854--7868, Aug. 2019.

\bibitem{Mezghani2018Blind}
A.~{Mezghani} and A.~L. {Swindlehurst}, ``Blind estimation of sparse broadband
  massive {MIMO} channels with ideal and one-bit {ADC}s,'' \emph{IEEE Trans.
  Signal Process.}, vol.~66, no.~11, pp. 2972--2983, June 2018.

\bibitem{kim2019channel}
I.~S. Kim and J.~Choi, ``Channel estimation via gradient pursuit for mm{W}ave
  massive {MIMO} systems with one-bit {ADC}s,'' \emph{EURASIP J. Wireless
  Commun. and Networking}, vol. 2019, no.~1, p. 289, 2019.

\bibitem{Mo2018Channel}
J.~Mo, P.~Schniter, and R.~W. Heath, ``Channel estimation in broadband
  millimeter wave {MIMO} systems with few-bit {ADC}s,'' \emph{IEEE Trans.
  Signal Process.}, vol.~66, no.~5, pp. 1141--1154, Mar. 2018.

\bibitem{Rodriguez2016Channel}
J.~Rodr{\'i}guez-Fern{\'a}ndez, N.~Gonz{\'a}lez-Prelcic, and R.~W. Heath,
  ``Channel estimation in mixed hybrid-low resolution {MIMO} architectures for
  mm{W}ave communication,'' in \emph{Proc. Asilomar Conf. Signals, Systems and
  Computers}, Pacific Grove, CA, USA, Nov. 2016, pp. 768--773.

\bibitem{Rusu2015Adaptive}
C.~Rusu, R.~Mendez-Rial, N.~Gonzalez-Prelcic, and R.~W. Heath, ``Adaptive
  one-bit compressive sensing with application to low-precision receivers at
  mm{W}ave,'' in \emph{Proc. IEEE Global Commun. Conf.}, San Diego, CA, USA,
  Dec. 2015.

\bibitem{Rao2019Channel}
S.~{Rao}, A.~{Mezghani}, and A.~L. {Swindlehurst}, ``Channel estimation in
  one-bit massive {MIMO} systems: {A}ngular versus unstructured models,''
  \emph{IEEE J. Select. Topics in Signal Process.}, vol.~13, no.~5, pp.
  1017--1031, Sep. 2019.

\bibitem{Ly2021SVM}
L.~V. Nguyen, A.~L. Swindlehurst, and D.~H.~N. Nguyen, ``{SVM}-based channel
  estimation and data detection for one-bit massive {MIMO} systems,''
  \emph{IEEE Trans. Signal Process.}, vol.~69, pp. 2086--2099, 2021.

\bibitem{balevi2019two}
E.~Balevi and J.~G. Andrews, ``Two-stage learning for uplink channel estimation
  in one-bit massive {MIMO},'' in \emph{Asilomar Conf. on Signals, Systems, and
  Computers}, Pacific Grove, CA, USA, Nov. 2019, pp. 1764--1768.

\bibitem{Dong2020Channel}
Y.~Dong, H.~Wang, and Y.-D. Yao, ``Channel estimation for one-bit multiuser
  massive {MIMO} using conditional {GAN},'' \emph{IEEE Commun. Letters},
  vol.~25, no.~3, pp. 854--858, Mar. 2021.

\bibitem{Zhang2020Deep}
Y.~{Zhang}, M.~{Alrabeiah}, and A.~{Alkhateeb}, ``Deep learning for massive
  {MIMO} with 1-bit {ADC}s: {W}hen more antennas need fewer pilots,''
  \emph{IEEE Wireless Commun. Letters}, vol.~9, no.~8, pp. 1273--1277, Aug.
  2020.

\bibitem{Kolomvakis2020Quantized}
N.~Kolomvakis, T.~Eriksson, M.~Coldrey, and M.~Viberg, ``Quantized uplink
  massive {MIMO} systems with linear receivers,'' in \emph{Proc. IEEE Int.
  Conf. Commun.}, Dublin, Ireland, June 2020.

\bibitem{DuyNguyen2020Neural}
D.~H.~N. {Nguyen}, ``Neural network-optimized channel estimator and training
  signal design for {MIMO} systems with few-bit {ADC}s,'' \emph{IEEE Signal
  Process. Letters}, vol.~27, pp. 1370--1374, 2020.

\bibitem{Gao2019Deep}
S.~{Gao}, P.~{Dong}, Z.~{Pan}, and G.~Y. {Li}, ``Deep learning based channel
  estimation for massive {MIMO} with mixed-resolution {ADC}s,'' \emph{IEEE
  Commun. Letters}, vol.~23, no.~11, pp. 1989--1993, Nov. 2019.

\bibitem{Zicheng2021Deep}
J.~Zicheng, G.~Shen, L.~Nan, P.~Zhiwen, and Y.~Xiaohu, ``Deep learning-based
  channel estimation for massive-{MIMO} with mixed-resolution {ADC}s and
  low-resolution information utilization,'' \emph{IEEE Access}, vol.~9, pp.
  54\,938--54\,950, Apr. 2021.

\bibitem{Jeon2018One}
Y.~Jeon, N.~Lee, S.~Hong, and R.~W. Heath, ``One-bit sphere decoding for uplink
  massive {MIMO} systems with one-bit {ADC}s,'' \emph{IEEE Trans. Wireless
  Commun.}, vol.~17, no.~7, pp. 4509--4521, July 2018.

\bibitem{Kim2020Machine}
S.~Kim, J.~Chae, and S.-N. Hong, ``Machine learning detectors for {MU-MIMO}
  systems with one-bit {ADCs},'' \emph{IEEE Access}, vol.~8, pp.
  86\,608--86\,616, Apr. 2020.

\bibitem{Lan2018Linearized}
A.~S. {Lan}, M.~{Chiang}, and C.~{Studer}, ``Linearized binary regression,'' in
  \emph{Proc. Annual Conf. on Inform. Sciences and Systems}, Princeton, NJ,
  USA, Mar. 2018.

\bibitem{Ly2021Linear}
L.~V. Nguyen, A.~Lee~Swindlehurst, and D.~H.~N. Nguyen, ``Linear and deep
  neural network-based receivers for massive {MIMO} systems with one-bit
  {ADC}s,'' \emph{IEEE Trans. Wireless Commun. (Early Access)}, 2021.

\bibitem{Demir2020ADMM}
O.~T. {Demir} and E.~{Bj\"{o}rnson}, ``{ADMM}-based one-bit quantized signal
  detection for massive {MIMO} systems with hardware impairments,'' in
  \emph{Proc. IEEE Int. Conf. on Acoustics, Speech and Signal Process.},
  Barcelona, Spain, May 2020, pp. 9120--9124.

\bibitem{Mirfarshbafan2020Algorithm}
S.~H. {Mirfarshbafan}, M.~{Shabany}, S.~A. {Nezamalhosseini}, and C.~{Studer},
  ``Algorithm and {VLSI} design for 1-bit data detection in massive
  {MIMO-OFDM},'' \emph{IEEE Open J. Circuits and Systems}, vol.~1, pp.
  170--184, Oct. 2020.

\bibitem{jeon2019robust}
Y.~{Jeon}, N.~{Lee}, and H.~V. {Poor}, ``Robust data detection for {MIMO}
  systems with one-bit {ADC}s: {A} reinforcement learning approach,''
  \emph{IEEE Trans. Wireless Commun.}, vol.~19, no.~3, pp. 1663--1676, Mar.
  2020.

\bibitem{Song2019CRC-Aided}
S.~H. {Song}, S.~{Lim}, G.~{Kwon}, and H.~{Park}, ``{CRC}-aided soft-output
  detection for uplink multi-user {MIMO} systems with one-bit {ADC}s,'' in
  \emph{Proc. IEEE Wireless Commun. and Networking Conf.}, Marrakesh, Morocco,
  Apr. 2019.

\bibitem{Cho2019OneBitSCSO}
Y.~{Cho} and S.~{Hong}, ``One-bit {S}uccessive-cancellation {S}oft-output
  ({OSS}) detector for uplink {MU-MIMO} systems with one-bit {ADC}s,''
  \emph{IEEE Access}, vol.~7, pp. 27\,172--27\,182, Feb. 2019.

\bibitem{Shao2018Iterative}
Z.~{Shao}, R.~C. {de Lamare}, and L.~T.~N. {Landau}, ``Iterative detection and
  decoding for large-scale multiple-antenna systems with 1-bit {ADC}s,''
  \emph{IEEE Wireless Commun. Letters}, vol.~7, no.~3, pp. 476--479, June 2018.

\bibitem{wen2016bayes}
C.~K. Wen, C.~J. Wang, S.~Jin, K.~K. Wong, and P.~Ting, ``{B}ayes-optimal joint
  channel-and-data estimation for massive {MIMO} with low-precision {ADC}s,''
  \emph{IEEE Trans. Signal Process.}, vol.~64, no.~10, pp. 2541--2556, May
  2016.

\bibitem{Thoota2021Variational}
S.~S. Thoota and C.~R. Murthy, ``Variational {B}ayes' joint channel estimation
  and soft symbol decoding for uplink massive {MIMO} systems with low
  resolution {ADC}s,'' \emph{IEEE Trans. Commun.}, vol.~69, no.~5, pp.
  3467--3481, May 2021.

\bibitem{Jeon2018supervised}
Y.~Jeon, S.~Hong, and N.~Lee, ``{S}upervised-learning-aided communication
  framework for {MIMO} systems with low-resolution {ADC}s,'' \emph{IEEE Trans.
  Veh. Technol.}, vol.~67, no.~8, pp. 7299--7313, Aug. 2018.

\bibitem{Ly2020Supervised}
L.~V. Nguyen, D.~T. Ngo, N.~H. Tran, A.~L. Swindlehurst, and D.~H.~N. Nguyen,
  ``Supervised and semi-supervised learning for {MIMO} blind detection with
  low-resolution {ADC}s,'' \emph{IEEE Trans. Wireless Commun.}, vol.~19, no.~4,
  pp. 2427--2442, Apr. 2020.

\bibitem{max1960quantizing}
J.~Max, ``Quantizing for minimum distortion,'' \emph{IRE Trans. Inf. Theory},
  vol.~6, no.~1, pp. 7--12, Mar. 1960.

\bibitem{pratt1981concavity}
J.~W. Pratt, ``Concavity of the log likelihood,'' \emph{J. the American
  Statistical Association}, vol.~76, no. 373, pp. 103--106, 1981.

\bibitem{bowling2009logistic}
S.~R. Bowling, M.~T. Khasawneh, S.~Kaewkuekool, and B.~R. Cho, ``A logistic
  approximation to the cumulative normal distribution,'' \emph{J. Industrial
  Engineering and Management}, vol.~2, no.~1, pp. 114--127, Mar. 2009.

\bibitem{Hershey-Unfolding-2014}
J.~R. Hershey, J.~L. Roux, and F.~Weninger, ``Deep unfolding: Model-based
  inspiration of novel deep architectures,'' \emph{arXiv:1409.2574}, 2014.

\bibitem{mezghani2012capacity}
A.~Mezghani and J.~A. Nossek, ``Capacity lower bound of {MIMO} channels with
  output quantization and correlated noise,'' in \emph{Proc. IEEE Int. Symp.
  Inform. Theory}, Cambridge, Massachusetts, USA, July 2012.

\bibitem{Nhan2020Deep}
N.~T. Nguyen and K.~Lee, ``Deep learning-aided {T}abu search detection for
  large {MIMO} systems,'' \emph{IEEE Trans. Wireless Commun.}, vol.~19, no.~6,
  pp. 4262--4275, June 2020.

\bibitem{Khani2020Adaptive}
M.~Khani, M.~Alizadeh, J.~Hoydis, and P.~Fleming, ``Adaptive neural signal
  detection for massive {MIMO},'' \emph{IEEE Trans. Wireless Commun.}, vol.~19,
  no.~8, pp. 5635--5648, Aug. 2020.

\bibitem{Mezghani2008Maximum}
A.~{Mezghani}, M.~{Khoufi}, and J.~A. {Nossek}, ``Maximum likelihood detection
  for quantized {MIMO} systems,'' in \emph{Proc. Int. ITG Workshop on Smart
  Antennas}, Vienna, Austria, Feb. 2008, pp. 278--284.

\bibitem{tensorflow}
\BIBentryALTinterwordspacing
M.~Abadi, A.~Agarwal, P.~Barham, E.~Brevdo, Z.~Chen, C.~Citro, G.~S. Corrado,
  A.~Davis, J.~Dean, M.~Devin, S.~Ghemawat, I.~Goodfellow, A.~Harp, G.~Irving,
  M.~Isard, Y.~Jia, R.~Jozefowicz, L.~Kaiser, M.~Kudlur, J.~Levenberg,
  D.~Man\'{e}, R.~Monga, S.~Moore, D.~Murray, C.~Olah, M.~Schuster, J.~Shlens,
  B.~Steiner, I.~Sutskever, K.~Talwar, P.~Tucker, V.~Vanhoucke, V.~Vasudevan,
  F.~Vi\'{e}gas, O.~Vinyals, P.~Warden, M.~Wattenberg, M.~Wicke, Y.~Yu, and
  X.~Zheng, ``{TensorFlow}: Large-scale machine learning on heterogeneous
  systems,'' 2015, {S}oftware available from tensorflow.org. [Online].
  Available: \url{https://www.tensorflow.org/}
\BIBentrySTDinterwordspacing

\bibitem{kingma2014adam}
D.~P. Kingma and J.~Ba, ``Adam: {A} method for stochastic optimization,''
  \emph{arXiv preprint arXiv:1412.6980}, 2014.

\end{thebibliography}


%\begin{IEEEbiography}{Author's Name}
%\end{IEEEbiography}

%\begin{IEEEbiographynophoto}{Author's Name}

%\newpage

%\begin{IEEEbiographynophoto}{Author's Name}


%\vfill


\end{document}